\documentclass{article}
\usepackage{arxiv}

\usepackage[utf8x]{inputenc} 
\usepackage[T1]{fontenc}
\usepackage{hyperref}
\usepackage{url}
\usepackage{microtype}  
\usepackage{graphicx}
\usepackage{epstopdf}
\usepackage{float}
\usepackage{amsmath}
\usepackage{amsfonts}
\usepackage{amssymb}
\usepackage{nccmath}
\usepackage{mwe}
\usepackage[table]{xcolor}
\usepackage{subcaption}
\usepackage{booktabs} 
\usepackage{nicefrac}  
\usepackage{rotating}
\usepackage{tikz}
\usetikzlibrary{tikzmark}

\usepackage[ruled,vlined]{algorithm2e}
\newcommand{\achal}[1]{{\color{black}{ #1}}}
\usepackage[toc,page]{appendix}
\usepackage{setspace}
\title{Fitting a stochastic model of intensive care occupancy to noisy hospitalization time series during the COVID-19 pandemic}

\author{Achal Awasthi$^{1}$, 
Volodymyr M. Minin$^{2}$, Jenny Huang$^{3}$, Daniel Chow$^{4}$, and  Jason Xu$^{1,3}$ \\
$^{1}$Department of Biostatistics and Bioinformatics, Duke University, U.S.A\\
$^{2}$Department of Statistics, University of California, Irvine, U.S.A\\
$^{3}$Department of Statistical Science, Duke University, U.S.A\\
$^{4}$School of Medicine, University of California, Irvine, U.S.A}

\date{}
\begin{document}
\maketitle

\begin{abstract}
Intensive care occupancy is an important indicator of health care stress that has been used to guide policy decisions during the COVID-19 pandemic. Toward reliable decision-making as a pandemic progresses, estimating the rates at which patients are admitted to and discharged from hospitals and intensive care units (ICUs) is crucial. Since individual-level hospital data are rarely available to modelers in each geographic locality of interest, it is important to develop tools for inferring these rates from publicly available daily numbers of hospital and ICU beds occupied. We develop such an estimation approach based on an immigration-death process that models fluctuations of ICU occupancy. Our flexible framework allows for immigration and death rates to depend on covariates, such as hospital bed occupancy and daily SARS-CoV-2 test positivity rate, which may drive changes in hospital ICU operations. We demonstrate via simulation studies that the proposed method performs well on noisy time series data and apply our statistical framework to hospitalization data from the University of California, Irvine (UCI) Health and Orange County, California. By introducing a likelihood-based framework where immigration and death rates can vary with covariates, we find, through rigorous model selection, that hospitalization and positivity rates are crucial covariates for modeling ICU stay dynamics and validate our per-patient ICU stay estimates using anonymized patient-level UCI hospital data.
\end{abstract}


\maketitle
\section{Introduction}
\label{s:intro}

In March 2020, the World Health Organization (WHO) declared the spread of a 2019 novel coronavirus (SARS-CoV-2) a global pandemic. Currently, we have seen over 350 million cases\footnote[2]{https://coronavirus.jhu.edu/map.html}, putting a tremendous strain on limited hospital capacities in countries around the world. Available hospital beds have proven to be one of the most valuable resources through the course of the pandemic, and in order to appropriately allocate resources, public health officials must rely on model forecasts to predict  occupancy during the current and future pandemics. \cite{ferstad2020model} Here, we propose a likelihood-based method for estimating the rates at which coronavirus disease 2019 (COVID-19) patients enter and leave hospital intensive care units (ICUs) based on publicly available hospitalization time series.\par

To quantify the uncertainties surrounding novel diseases such as COVID-19, constant policy changes, and changing rates of vaccine distribution, stochastic epidemic models are well-suited compared to their deterministic counterparts. However, stochastic modeling of SARS-CoV-2 transmission dynamics has not been more widely adopted, in large part due to computational burdens associated with inference for stochastic model classes such as partially observed Markov processes. \cite{ionides2006inference,breto2018modeling} COVID-19 observational data are often incomplete, marked with coarseness and systematic under-reporting. This often gives rise to intractable marginal likelihood functions associated with partially observed data. Although likelihood-based methods provide principled ways to estimate parameters and quantify uncertainty, fitting a stochastic model in such missing data settings requires marginalizing over the set of all possible events that are compatible with the observed data. When this latent space is enormous, fitting stochastic epidemic models becomes a highly non-trivial exercise due to the increased computational complexity. \cite{doss2013fitting, xu2015likelihood} This has spurred recent work on numerical approaches and approximation methods toward likelihood-based inference. \cite{crawford2012transition, fearnhead2014inference, ho2018birth, de2020parameter,  FintziWakefieldMinin} \par

In this article, we overcome this intractability for a class of immigration-death processes representing ICU occupancy. We focus our attention on the simpler problem of estimating the dynamics and mean duration of ICU stays instead of the full dynamics of SARS-CoV-2 transmission through a population. Our approach accounts for the missing data probabilistically in a principled way by directly utilizing the likelihood of the observed data. We present a methodological framework for maximum likelihood estimation for a class of continuous time immigration-death processes from population-level data. Our models fall within the class of continuous-time Markov chains, and we carefully balance their flexibility with tractability to allow for deriving expressions for the transition probabilities of the process. These comprise the marginal likelihood function; their solutions together with numerical and optimization techniques enable classical procedures such as maximum likelihood estimation. As a result, we can perform principled inference of interpretable model parameters in a mechanistic model, integral to understanding the mean duration of ICU stay by COVID-19 patients according to the underlying dynamic of the pandemic.\par

Specifically, we begin by assuming that the per-patient immigration rate, i.e., the rate at which patients are admitted to the ICU, is proportional to the daily number of hospital beds occupied, while the per-patient death or \achal{clearance} rate, i.e., the rate at which patients are discharged from the ICU, is constant over time. As previous studies have shown that the length of hospital ICU stay varies depending on resource availability, patient characteristics, and hospital-specific differences in clinical care strategies,\cite{lescure2021sarilumab} we extend the methodology to allow the model parameters to depend on a number of covariates. The proposed framework is flexible as a result, allowing immigration and\achal{clearance} rates to vary with a number of salient variables in a log-linear fashion. \par

The rest of the paper is structured as follows. In section \ref{Data and the Immigration-Death Process}, we present a description of the data that motivated us and also describe an immigration-death process in detail. In section \ref{Covariate-dependent rate parameterization}, we introduce the the parametrization of the immigration rates based on relevant covariates and present our immigration-death framework in detail. In section \ref{Fitting an immigration-death model}, we present derivations of the Kolmogorov forward equations, transition probabilities, and the likelihood function for discretely observed data from an immigration-death process. We also briefly describe a technique that we use to calculate the transition probabilities in a computationally efficient way. In section \ref{Optimizing the observed likelihood function}, we describe the optimization methods used to numerically calculate the MLE and uncertainty quantification for the MLE. \par

In section \ref{Simulation study}, we design a simulation study to test our immigration-death framework. We then apply our model to hospitalization data from the UC Irvine hospital system as well as Orange County in sections \ref{University of California, Irvine} and \ref{Application to COVID-19 hospitalization data from Orange County, CA} respectively, exploring the division of data into three phases following natural waves of the pandemic. The concluding section \ref{s:discuss} discusses limitations and interpretations of the model as well as directions towards future work.

\section{Methods}
\label{sec2:Methods}

We begin by describing the data that motivate the immigration-death model. In our application to the problem of ICU stay estimation, the immigration and death rates will be interpretable as the rates at which patients are admitted to and discharged from the ICU respectively. In order to provide enough realism in the model without overfitting, we test and select among models of varying complexity. These methods hinge on access to the observed data likelihood, which additionally yields a natural approach to uncertainty quantification of the parameter estimates.

\subsection{Data and the Immigration-Death Process}
\label{Data and the Immigration-Death Process}
`We consider hospital-level data collected from March 2020 to December 2020, consisting of daily counts of occupied hospital beds and ICU beds in the University of California, Irvine's hospital system (UCI Health) and the larger Orange County (OC) level. \achal{There are $27$ hospitals in Orange County (OC) according to the OC health Care Agency; out of these $27$ hospitals, $25$ of them can serve as Emergency Receiving Centers. The number of hospital beds and ICU beds occupied daily in OC also referred to as hospitalization data for OC is publicly available\footnote{https://occovid19.ochealthinfo.com/coronavirus-in-oc} consisting of county-level surveillance data that are obtained by aggregating the individual numbers of hospital beds and ICU beds occupied daily as reported by each major hospital in the county.} We denote the aggregated number of hospital beds and ICU beds occupied by COVID-19 patients in OC by $H_{0}, H_{1}, \ldots H_{T}$ and $I_{0}, I_{1}, \ldots I_{T}$ respectively. \achal{Although data were reported on a daily basis, on any given day in the observation period, it was typical for 2-5 hospitals to miss out on reporting. This underreporting may have arisen from reasons not limited to being short-staffed, neglecting to report on weekends, or not reporting on time. On the contrary, patient-level data also referred to as ``line-list'' data obtained from the UCI Medical Center include the average monthly durations of ICU stay for all patients admitted to the UCI Medical Center ICU with a confirmed diagnosis of COVID-19.}\par

\begin{figure}[h]
\centering
\captionsetup[subfigure]{oneside,margin={-2cm,0cm}}
\begin{subfigure}{0.50\textwidth}
   \includegraphics[width=0.75\linewidth]{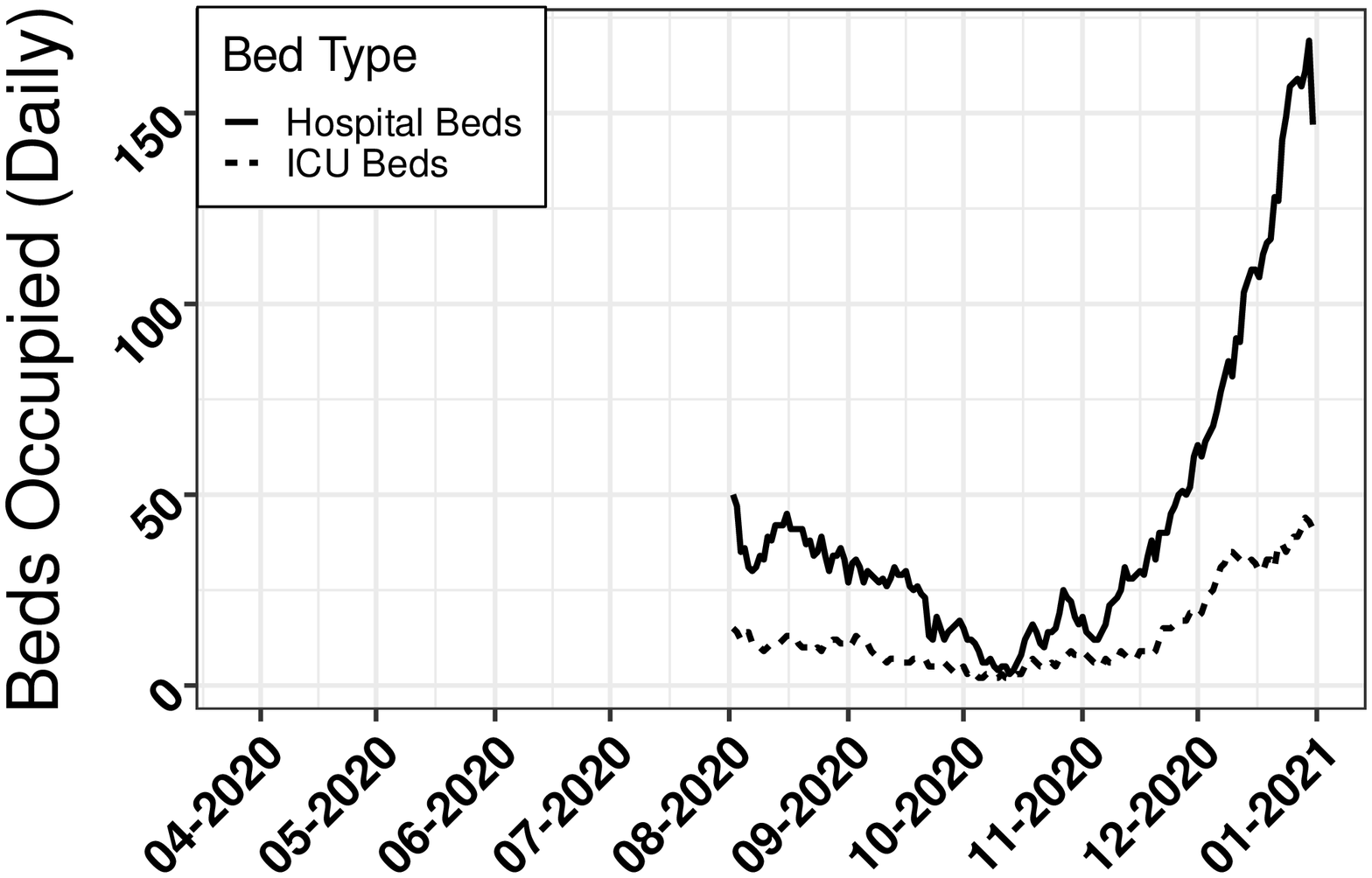}
   \vspace{-0.75cm}
   \caption{}
   \label{fig4:sub1}
\end{subfigure}
\begin{subfigure}{0.50\textwidth}
   \includegraphics[width=0.75\linewidth]{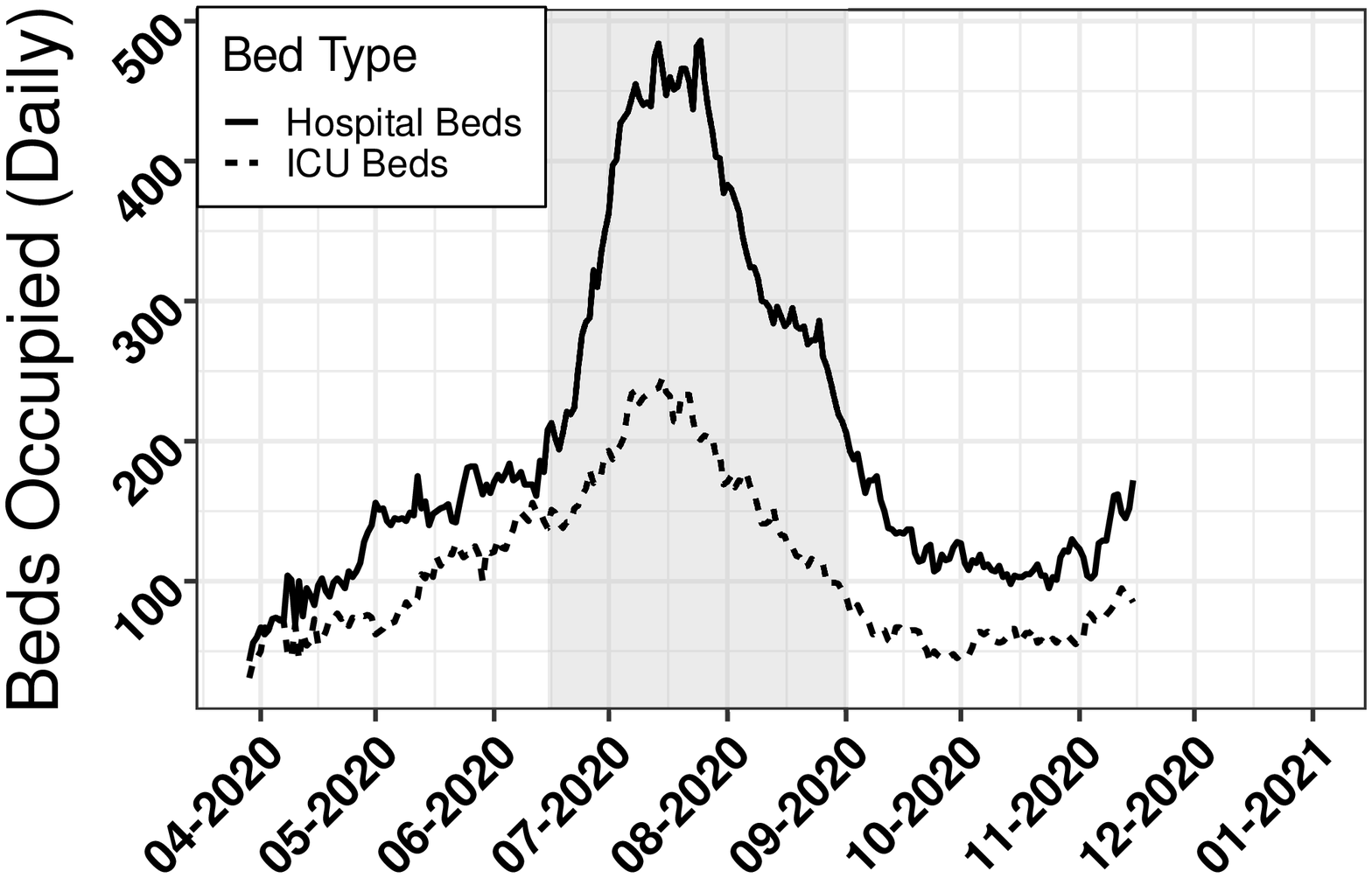}
   \vspace{-0.75cm}
   \caption{}
   \label{fig4:sub2}
\end{subfigure}
\begin{subfigure}{0.50\textwidth}
   \includegraphics[width=0.75\linewidth]{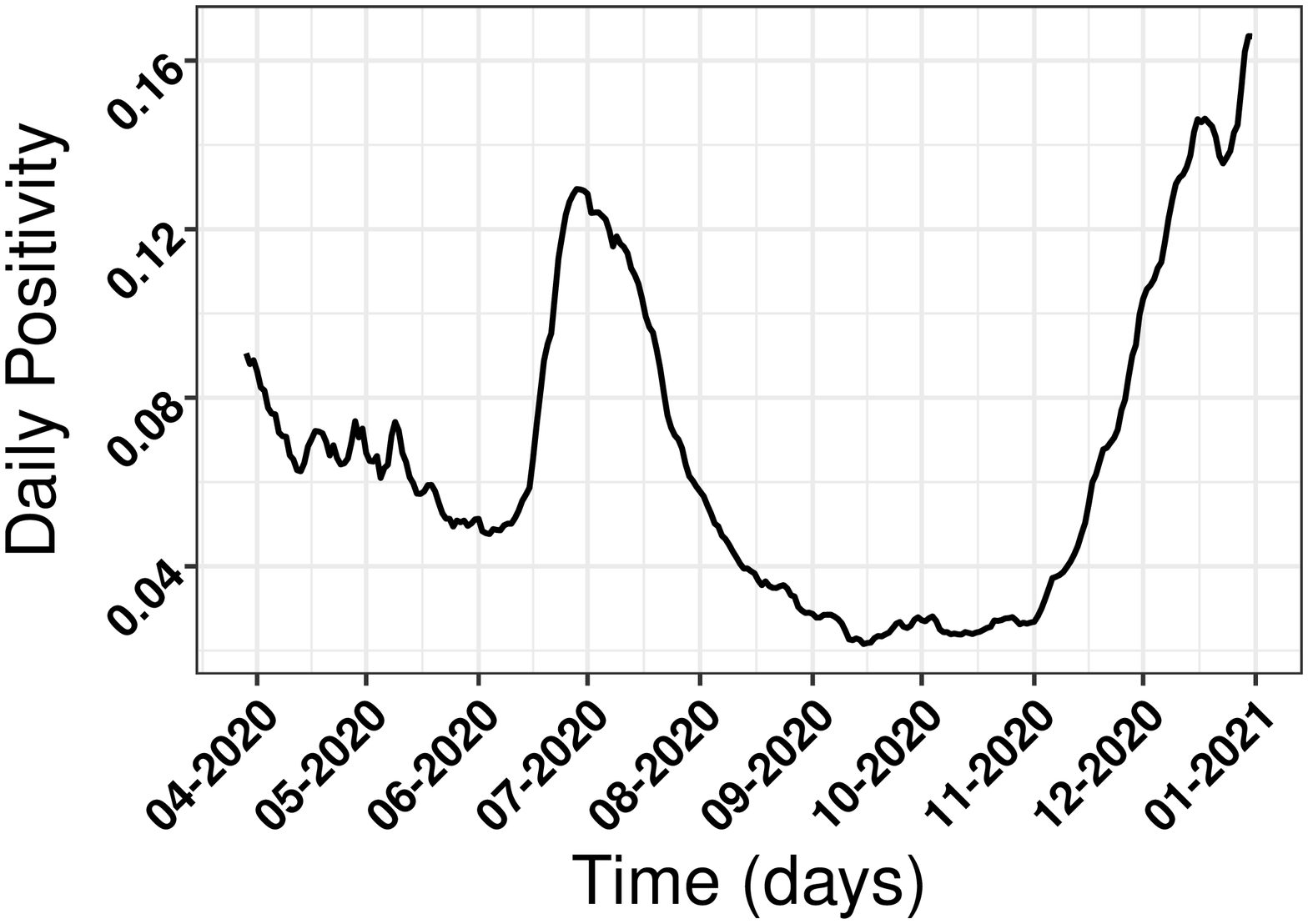}
   \vspace{-0.25cm}
   \caption{}
   \label{fig4:sub3}
\end{subfigure}
\caption{a) Daily hospital and ICU bed counts for UCI Health between August 02, 2020 and December 31, 2020. b) Daily hospital and ICU bed counts for OC between March 29, 2020 and November 15, 2020 divided into three phases. (c) Daily cumulative test positivity rate for OC.}
\label{fig:4}
\end{figure}

Figures \ref{fig4:sub1} and \ref{fig4:sub2} represent these data in the UCI hospital system and in all of Orange County, respectively. For the rest of the paper, hospitalization data refers to both the number of hospital beds occupied and the number of ICU beds occupied.

Our goal is to fit a stochastic immigration-death model of ICU occupancy to these data, inferring the per-patient rates of ICU admission as well as the rate of clearance. An immigration-death process $X(t)$ is a continuous-time Markov chain (CTMC) defined over a countable state space of the non-negative integers. The process can increase by one in the case of an immigration event, decrease by one in the event of a death (clearance), or remain at the same state at any instant in time $t>0$. The process $X(t)$ can be completely characterized by its immigration rate $\alpha(t)$ and its death rate $i \mu(t)$---we model the latter as proportional to the current state of the process, making the death process rate-linear. If the process is in state $X(t)=i$ at time $t$,     
then the process can be defined by its instantaneous rates $\lambda_{ij}(t)$ which parametrize the infinitesimal generator matrix  $\pmb{\Lambda} (t) = \{\lambda_{ij}(t) \}$ characterizing the time-inhomogeneous CTMC. To understand the relation of instantaneous rates to the behavior of the process, denote the \textit{transition probabilities}  $\mathbb{P}_{ij}(s) = \mathbb{P}(X(t+s)=j|X(t)=i)$, defined as the probability of the process transitioning from state $i$ at time $t$ to state $j$ at time $t+s$.  Then for a small time interval $\Delta t$, the probabilities of an immigration event, a death event, and no event occurring are given by

\begin{equation}
    \mathbb{P}_{ij}(\Delta t)=%
    \begin{cases} 
      \alpha(t)\Delta t + o(\Delta t) & j=i+1, \\
      i\mu(t) \Delta t +  o(\Delta t) & j=i-1, \\
      1 - (\alpha(t)\Delta t + i\mu(t) \Delta t + o(\Delta t)) & j=i, \\
      0 & \text{otherwise}.
   \end{cases}
   \label{eq:transprob}
\end{equation}

Since we model ICU occupancy as the immigration-death process $X(t)$, naturally immigration can be interpreted as the flow of patients into the ICU, while death is interpreted as a patient leaving the ICU when they are discharged or they die.\achal{However, in order to disambiguate the term ``death'' for our application to the problem of ICU stay, we refer to the ``death'' rate of a stochastic immigration-death model as ``clearance'' rate throughout this paper.} Therefore, we view the count data $\{ I_0, I_1, \ldots I_T \}$ as discrete observations of the continuous-time process, that is, $I_1 = X(t_1)$ for a sequence of observation times $t_i$ at which the counts are available. The immigration rate $\alpha(t)$ will depend on hospitalization and other factors, described in the next subsection, while the per-person\achal{clearance rate} $\mu(t) = \mu$ is assumed to be homogeneous throughout the observation period $[0,T]$. The summary of the model is illustrated in Figure 1 appearing in the supporting information. Note that the overall rate of ICU clearance $i \mu$ fluctuates proportionally with the current ICU occupancy, a natural modeling assumption in our context.

\subsection{Covariate-dependent rate parameterization}
\label{Covariate-dependent rate parameterization}

As the model described above may lack realism in positing that the immigration and (per-patient) clearance rates are constant throughout the entire observation period, we add flexibility by allowing them to depend on potentially time-varying covariates. We parametrize the log immigration rate as a linear combination of salient covariates while assuming a constant clearance rate over the entire observation period. Equation \ref{eq:rates_covariates} describes such an immigration rate where $\log\alpha$ plays the role of an intercept term,\achal{$\psi_{i}(t)$} represents an $i^{th}$ covariate of interest at time $t$, and $\beta_{i}$ is its corresponding coefficient: 

\begin{equation}
   \log \alpha(t) = \log\alpha + \sum_{i=1}^{k}\beta_{i} \achal{\psi_{i}(t)} , \quad \mu(t) = \mu.
   \label{eq:rates_covariates}
\end{equation}

This parametrization enables us to account for a set of covariates that may be relevant in accurately describing how the model rates vary. As a large inflow of patients to the ICU comes from patients already admitted in the hospital, the number of patients admitted to the hospital is the leading predictor affecting the immigration rate in the above parametrization. We hypothesize that the proportion of infected people in a particular region is also likely to influence the rate of patients entering the ICUs. Thus, an overwhelming number of patients already admitted to the hospital combined with a higher proportion of infected people in a particular region suggest that the patients will enter the ICU at a higher rate. In particular, we allow the per-patient immigration rate into the ICU to depend on factors such as the daily number of hospital beds occupied $H_{O}(t)$, and the daily cumulative test positivity rate $P_A(t)$ in the region of interest. \par

The daily cumulative test positivity rate $P_A(t)$ is defined as the ratio of the cumulative number of individuals who tested positive for COVID-19 in the last seven days to the number of individuals who were tested for COVID-19 in the last seven days. This daily cumulative positivity rate serves as an indicator of case load on a particular day, and takes values between $0$ and $1$, with a higher value of $P_{A}(t)$ indicates that a greater proportion of people have tested positive for COVID-19. Higher positivity rates in a region may lead to changes in the decision-making process within hospital systems as there are a limited number of hospital and ICU beds available on a particular day. 

\begin{table}[h]
\centering
\begin{tabular}{c l c}
\textbf{Model} & \textbf{Immigration Rate}  & \# \textbf{Parameters} \\ \hline
 1 & $\log{\alpha(t)} = \log{\alpha} + \beta_{H} \log{H_{O}(t)} + \beta_{H:P}P_{A}(t)\log{H_{O}(t)} + \beta_{P}P_{A}(t)$  & 5\\ 
 2 & $\log{\alpha(t)} = \log{\alpha} + \beta_{H} \log{H_{O}(t)} + \beta_{H:P}P_{A}(t)\log{H_{O}(t)}$  & 4 \\ 
 3 & $\log{\alpha(t)} = \log{\alpha} + \beta_{H} \log{H_{O}(t)} + \beta_{P}P_{A}(t)$  & 4 \\ 
 4 & $\log{\alpha(t)} = \log{\alpha} + \beta_{H} \log{H_{O}(t)}$  & 3 \\
 5 & $\log{\alpha(t)} = \log{\alpha} + \log{H_{O}(t)}$  & 2 \\ \hline
\end{tabular}
\vspace{2mm}
\caption{Various models considered that differ by the way they parametrize (log) immigration rates}
\label{tab1}
\end{table}

Though it is intuitive that the potential covariates we consider would impact the rates of the process, it is preferable to quantitatively select the features to include in a way that balances model flexibility and generalizability given limited data. Within a likelihood-based framework, information-theoretic  criteria allow for rigorous model selection. For instance, we may select covariates while guarding against overfitting using the Akaike information criterion (AIC)
$$-2l(\achal{\mathbf{X}};\mathbf{\Theta}) + 2K,$$ where $K$ is the number of free parameters and $l(\achal{\mathbf{X}};\mathbf{\Theta})$ is the maximized log-likelihood of the given data. Among a set of candidate models, the one with lowest AIC is preferred. We explore several competing models that posit different log-linear functional forms for $\alpha(t)$, summarized in Table \ref{tab1}, where $\alpha, \mu, \beta_{H:P}$, and $\beta_{H}$ are non-negative coefficients. 

\subsection{Fitting an immigration-death model}
\label{Fitting an immigration-death model}
Toward enabling likelihood-based inference, this subsection derives the transition probabilities of a simple immigration-death process, describes how the transition probabilities are related to the marginal likelihood function, and describes a technique to obtain the transition probabilities in a computationally efficient way. \par 

Recall the transition probability $\mathbb{P}_{ij}(s) = \mathbb{P}(X(t+s)=j|X(t)=i)$ denotes the probability of the immigration-death process transitioning from state $i$ at time $t$ to state $j$ at time $t+s$. Evaluating the likelihood in the proposed model framework relies on  access to these quantities as they appear as a product in the likelihood function of the observed data. We outline a method that derives their solution by way of the probability generating function (PGF), and also describe an efficient numerical way to compute them. This will allow us to numerically optimize the likelihood function toward parameter estimation. Similarly, we then detail how to perform interval estimation and covariate-based model selection efficiently using these numerical techniques. \par

Let $\{X(1), X(2) \dots X(T)\}$ denote the  numbers of occupied ICU beds at observation times $t_1,t_2,\ldots t_T$. By the Markov property, the likelihood function of these discretely observed data is given by a product of the transition probabilities

\begin{equation}
    \mathcal{L(\achal{\mathbf{X}};\mathbf{\Theta})} = \prod_{i=1}^{T-1}\mathbb{P}_{X(i)X(i+1)}(t_{i+1} - t_{i}),
    \label{eq:lik}
\end{equation}

where $\mathbf{\Theta}$ denotes the set of parameters we wish to infer. Thus, it is necessary to calculate the transition probabilities as they form the backbone of the observed likelihood function. Next we derive the transition probabilities for a discretely observed immigration-death process with immigration rate $\alpha(t)$ and death or clearance rate $\mu(t)$. \par
We begin by deriving the Kolmogorov's equations \cite{KARLIN19751} for a discretely observed immigration-death process:
\begin{align*}
    \mathbb{P}_{ij}(t+h) &= \sum_{k}\mathbb{P}_{ik}(t)\mathbb{P}_{kj}(h) \\
    &= \mathbb{P}_{ij-1}(t)\mathbb{P}_{j-1,j}(h) + \mathbb{P}_{ij}(t)\mathbb{P}_{jj}(h) + \mathbb{P}_{ij+1}(t)\mathbb{P}_{j+1,j}(h) + \sum_{k\neq j,j-1,j+1}\mathbb{P}_{ik}(t)\mathbb{P}_{kj}(h) \\
    &= \mathbb{P}_{ij-1}(t)(\alpha(t)h + o(h)) + \mathbb{P}_{ij}(t)(1 - \alpha(t)h - j\mu(t)h - o(h))\\ 
    &+ \mathbb{P}_{ij+1}(t)((j+1)\mu(t)h+o(h))+ o(h) \\
    &= \mathbb{P}_{ij-1}(t)\alpha(t)h + \mathbb{P}_{ij-1}(t)o(h) + \mathbb{P}_{ij}(t) - \mathbb{P}_{ij}(t)\alpha(t)h - \mathbb{P}_{ij}(t)j\mu(t)h - \mathbb{P}_{ij}(t)o(h)\\ 
    &+ \mathbb{P}_{ij+1}(t) (j+1)\mu(t)h + \mathbb{P}_{ij+1}(t)o(h)+ o(h),\\
\end{align*}
where the third equality is due to Equation \ref{eq:transprob}. After subtracting $\mathbb{P}_{ij}(t)$ from both the sides, dividing by h, and sending $h \to 0$, we obtain the Kolmogorov equations:
\begin{align*}
    \mathbb{P'}_{ij}(t) &= (-\alpha(t) - j\mu(t))\mathbb{P}_{ij}(t) + \alpha(t)\mathbb{P}_{ij-1}(t)
    +(j+1) \mu(t) \mathbb{P}_{ij+1}(t),\\
    \mathbb{P'}_{i0}(t) &= -\alpha(t)\mathbb{P}_{i0}(t) + \mu(t) \mathbb{P}_{i1}(t).
    \label{eq1}
\end{align*}
\achal{We will use the Kolmogorov's equations together with Equation \ref{eq2} to derive the closed form solution for the PGF $\phi_{i}(s,t)$.} Next we consider the PGF,
\begin{equation}
    \phi_{i}(s,t) := \mathbb{E}[s^{X(t)}|X(0)=i] = \sum_{j=0}^{\infty}\mathbb{P}_{ij}(t) s^{j}
    \label{eq2}
\end{equation}
for $s \in [0,1]$. We differentiate Equation (\ref{eq2}) with respect to time and use Kolmogorov's equations to yield Kendall’s partial differential equation (PDE), \cite{KARLIN19751, lange2003applied}
\begin{align*}
    \frac{\partial}{\partial{t}}\phi_{i}(s,t)
    = \Bigg[(1-s)\mu(t) \frac{\partial}{\partial{s}} + (s-1)\alpha(t)\Bigg]\phi_{i}(s,t),
\end{align*}
subject to the initial condition $\phi_{i}(s,0) = s^{i}$. A detailed derivation of the steps toward obtaining and solving Kendall's PDE is available in Appendix \ref{Appendix_1}.
The solution of the above PDE is
\begin{equation}
    \phi_{i}(s,t) = \exp\Bigg[{{\dfrac{\alpha(t)(1-s)\Big(e^{-\mu(t) t} - 1\Big)}{\mu(t)}}}\Bigg]\Bigg[1+(s-1)e^{-\mu(t) t}\Bigg]^{i}\achal{.}
    \label{eq:PGF}
\end{equation}
\achal{This closed form solution for the PGF aids in the calculation of transition probabilities.} Notice the transition probability $\mathbb{P}_{ij}(t) = \mathbb{P}(X(t)=j|X(0)=i)$ can formally be obtained by differentiating the PGF $\phi_{i}(s,t)$ and normalizing by an appropriate constant,
\begin{equation}
    \mathbb{P}_{ij}(t) = \frac{1}{j!}\frac{\partial^j}{\partial{s}}\phi_{i}(s,t)\Bigr|_{\substack{s=0}}.
    \label{eq5}
\end{equation}
However, in practice repeated numerical differentiation is a computationally intensive procedure and often becomes numerically unstable, especially for a large $j$. Instead, we make use of the Fast Fourier transform (FFT) to calculate the transition probabilities $P_{ij}(t)$ which occur as coefficients of $s^{j}$ in Equation (\ref{eq2}).\cite{lange1982calculation}\par

We map the domain $s\in [0,1]$ onto the boundary of the complex unit circle, considering a change of variables $s=e^{2\pi i\omega}$, so that the new domain becomes an equally spaced set of points along the unit circle in the complex plane. This allows us to view the generating function defined in Equation (\ref{eq2}) as a periodic function
$$\phi_{i}(e^{2\pi i\omega},t) = \sum_{j=0}^{\infty} P_{ij}(t)e^{2\pi i\omega j},$$
where the coefficients of interest $P_{ij}$ are now interpretable as the $j^{th}$ Fourier coefficients of a Fourier series. This suggests they can be recovered by the Fourier inversion formula,
$$P_{ij}(t) = \int_{0}^{1} \phi_{i}(e^{2\pi i\omega},t)e^{-2\pi i\omega j} d\omega.$$ 

A Riemann sum approximation can be used to discretize the integral,
\begin{equation}
    P_{ij}(t) \approx \frac{1}{N-1}\sum_{k=0}^{N-1}  \phi_{i}(e^{2\pi i k/N},t)e^{-2\pi i\omega j/N},
    \label{eq:rsum}
\end{equation}
with a larger choice of N yielding a more precise approximation. This approach avoids the need to directly compute numerical higher-order partial derivatives of $\phi_{i}(s,t)$ as described in Equation (\ref{eq5}). Instead, the FFT provides a fast way to extract \textit{all} of the coefficients $P_{ij}(t)$  simultaneously for $ 0 \leq j \leq N-1$. \par

Recall from Equation \eqref{eq:lik} that the likelihood function of the discretely observed data is comprised of these transition probabilities. This FFT method therefore enables its efficient computation for any given set of parameter values, which can then be embedded within any generic optimization scheme that requires calls to the likelihood $\mathcal{L(\achal{\mathbf{X}};\mathbf{\Theta})}$ as the objective function. 

\subsection{Optimizing the observed likelihood function} 
\label{Optimizing the observed likelihood function}
 
With an efficient way to compute transition probabilities in hand, we can use Equation (\ref{eq:lik}) to numerically evaluate the  likelihood function $\mathcal{L(\achal{\mathbf{X}};\mathbf{\Theta})}$ for any  parameter configuration, and in turn use generic iterative optimization methods to maximize it.\cite{xu2015likelihood} We find that the Nelder-Mead method \cite{NeldMead65} delivers robust yet efficient solutions to maximizing the log-likelihood, by way of a simplex algorithm using only calls to the objective function. That is, the method does not require gradient or other higher-order information.\cite{nocedal2006numerical} Because we cannot guarantee the convexity of our objective function, the algorithm is only guaranteed to converge to a local optimum.\cite{nocedal2006numerical} As a result, we advocate initializing the algorithm using multiple restarts from different initial starting values. \cite{kochenderfer2019algorithms} We find in our empirical studies that numerically optimizing the log-likelihood using the Nelder-Mead algorithm is stable and can repeatedly deliver a consistent optimal solution from several distinct initializations. \par

With a robust method for computing the MLE, we can similarly obtain interval estimates numerically.With a robust method for computing the MLE, we can similarly obtain interval estimates numerically.\achal{Recall that asymptotic normality of the maximum likelihood estimator can be used toward constructing the confidence intervals numerically. Denoting $\pmb{\hat{\theta}_{ML}}$ as the maximum likelihood estimate for $\pmb{\theta}$, Fisher's classical approximation theorem establishes that its asymptotic distribution is given by
\begin{align*}
    \pmb{\hat{\theta}_{ML}} \sim \mathcal{N}(\pmb{\theta}, [I(\pmb{\theta})]^{-1}_{n \times n}),
\end{align*}
under mild assumptions, where $I[(\pmb{\theta})]_{n \times n}$ is the Fisher information matrix. These results also hold for maximum likelihood estimates under Markov chain models \cite{billingsley1961statistical}, while Cronie and Yu (2016)  derive a version catered to evenly spaced observations from an immigration-death process, coinciding with our baseline model of ICU stays \textit{X(t)}. In all cases, these asymptotic results entail approximate intervals. We find that applying Fisher's approximation stated above performs well empirically in all cases we consider here; it is appealing in its simplicity, especially in settings with covariate-parametrized rates.}
In theory, standard errors for parameter estimates are given upon inversion, though the information depends on the unknown parameter values. In practice, we calculate the observed Fisher information $[I(\pmb{\hat{\theta}_{ML}})]_{n \times n}$ evaluated at the maximum likelihood estimates. That is, we obtain the Hessian of the log-likelihood evaluated at the MLE using numerical differentiation at convergence. 
Upon inversion, using its diagonal entries allows us to construct approximate $95\%$ confidence intervals, given by
$$
\hat{\theta}_{i} \pm 1.96 \sqrt{a_{ii}}, \quad i=1,2, \ldots n,
$$
where $A = [I(\pmb{\hat{\theta}_{ML}})]_{n \times n}^{-1}$ and $a_{ij}$ denotes its $(i,j)^{\text{th}}$ entry. 

\section{Empirical Performance and Results}
\label{sec3:res}

\subsection{Simulation study}
\label{Simulation study}
We now assess the proposed methodology via simulation. The experiment aims to recover the true parameters used to generate the synthetic data from the immigration-death model given data at only a set of discrete observation times. We design the simulations to resemble the real data in one of our case studies, with the ground truth parameters and observation schedule being chosen so that the shape and scale of the simulated outbreaks reflects that of the OC hospitalization data we seek to analyze. \par

The daily number of hospital and ICU beds occupied in the county-level OC data is obtained by aggregating the individual numbers of hospital and ICU beds occupied as reported by each major hospital in the county. We replicate this in our simulations by first generating the number of hospital and ICU beds occupied for several hospitals, and then summing across hospitals to obtain population-level hospitalization data that is comparable to the OC data we will study. As it is possible that some hospitals do not report hospitalization data on a particular day,  we additionally extend our simulation study to assess the robustness and potential bias of the proposed framework in the presence of underreporting. Note that we might expect the simulated hospitalization data for each hospital in the absence of any underreporting to resemble the UCI Health hospitalization data, while the aggregated hospitalization data in the presence of underreporting should resemble the noisier hospitalization data at the county level.

\paragraph{{Simulation Procedure}}
\label{sim_procedure}
To establish notation, let $\tilde{H}_{tj}$ and $\tilde{I}_{tj}$ be the number of hospital beds and ICU beds occupied on the $t^{th}$ day in the $j^{th}$ hospital and $[0, 1, 2, \ldots T_{max}]$ be the discrete times at which we observe the process with $[0, T_{max}]$ being the total observation period. We first generate the daily number of hospital beds occupied for each of the $m$ hospitals. To reflect scale of the population-level data in our case study, we simulate individual hospital occupancies so that the total number of occupied beds across hospitals matches our dataset. To this end, we begin by sampling $\tilde{H}_{0j}$ from a multinomial distribution such that $\sum_{j=1}^{m} \tilde{H}_{0j} = H_{O}(0)$, the total number of hospital beds at the start of our observation period in the OC data. For each day in the observation period, we next sample the difference $d_{tj} = \tilde{H}_{tj} -\tilde{H}_{(t-1)j}$ multinomially with equal weights such that $\sum_{j=1}^{m} d_{tj} = H(t) - H(t-1), 1 \leq t \leq T_{max}$. The daily number of hospital beds occupied, $\tilde{H}(t)$, during the entire observation period for $m$ hospitals is then obtained by cumulatively summing $d_{tj}$. \par

Conditional on the simulated hospital bed counts  $\tilde{H}_{tj}$, we generate the daily numbers of ICU beds occupied from an immigration-death process with rates 

\begin{equation}
     \log \alpha(t) = \log \alpha + \log \tilde{H}(t),  \quad \mu(t) = \mu, \quad 0 \leq t \leq T_{max}. \vspace{-1pt}
     \label{eqsim:rates}
\end{equation} 

We use the Gillespie algorithm \cite{gillespie1977exact} to generate realizations of the process defined by the rates in Equation (\ref{eqsim:rates}), summarized by Algorithm 1 appearing in the supporting information. We hold out information from these trajectories between observation times, summarizing only the daily counts of the simulated Markov process to get $\tilde{I}_{tj}, \mkern3mu 0 \leq t \leq T_{max}, \mkern3mu 0\leq j \leq m$, thus resulting in a time-series of number of hospital beds occupied $\mathbf{\tilde{H}}$ and ICU beds occupied $\mathbf{\tilde{I}}$ :
\\
\\
\[
\mathbf{\tilde{H}} =
\tikzmarknode{mat}
  {\begin{bmatrix}
    H_{01} & H_{12} & \ldots & H_{1m} \\
    H_{11} & H_{22} & \ldots & H_{2m} \\
    \ldots &  \ldots &  \ldots &  \ldots & \\
    H_{T_{max}1} & H_{T_{max}2} & \ldots & H_{T_{max}m}
  \end{bmatrix}},
\begin{tikzpicture}[overlay,remember picture]
\draw[blue,thick,-latex] node[anchor=south west] (nn1) at (mat.north west)
{number of hospitals} (nn1.east) -- (nn1-|mat.north east) 
node[midway,above,black]{};
\draw[red,thick,-latex] node[anchor=north east,align=center] (nn2) at (mat.north west)
{number of days} (nn2.south) -- (nn2.south|-mat.south west) 
node[midway,above,black,rotate=90]{};
\end{tikzpicture}
  \qquad
 \mathbf{\tilde{I}} =
  \begin{bmatrix}
    I_{01} & I_{12} & \ldots & I_{1m} \\
    I_{11} & I_{22} & \ldots & I_{2m} \\
    \ldots &  \ldots &  \ldots &  \ldots & \\
    I_{T_{max}1} & I_{T_{max}2} & \ldots & I_{T_{max}m}
  \end{bmatrix}  
  .\]
 \\
In all simulations, the input consists of the true parameter values $\mathbf{\Theta}= \{\alpha, \mu\} = \{0.1, 0.2\}$. To generate realizations comparable to the real dataset, we choose
$H(0) = 43$, $m = 25$,   $T_{max} = 200$ days, $\tilde{I}_{0j} = 1$ for all $j=1, \ldots, 25$, and  number of trials $n_{sim} = 1000$.\par 

After independently simulating $1000$ instances of $\tilde{H}_{tj}$,  we simulate $\tilde{I}_{tj}$ conditional on each $\tilde{H}_{tj}$ as described above. We then aggregate each of the $1000$ instances of $\tilde{H}_{tj}$ and $\tilde{I}_{tj}$, 
\begin{equation}\label{eq:cum}
    \tilde{H}(t) = \sum_{j=1}^{25} \tilde{H}_{tj}, \quad \tilde{I}(t)=\sum_{j=1}^{25} \tilde{I}_{tj}, \quad 0 \leq t \leq 200,
\end{equation} to obtain the  daily total hospital and ICU beds occupied respectively. \par

In order to capture possible underreporting by hospitals, we next consider a misspecified simulation by randomly sampling a number $\gamma_{t} \sim Unif\{0,1,2\}$ of hospitals to be ``dropped'' on each day in the observation period. These dropped hospitals are chosen uniformly at random from the 25 hospitals, and are omitted from the sums \eqref{eq:cum} before aggregating the counts $\tilde{H}_{tj}$ and $\tilde{I}_{tj}$. \par

\paragraph{Results} 
For both studies with or without underreporting, we calculate the transition probabilities using Equation (\ref{eq:rsum}) with $N=2^{11}$, and numerically optimize the log-likelihood from ten random restarts under the Nelder-Mead algorithm. The average MLE across the $1000$ simulations for $\mathbf{\Theta} = \{\alpha, \mu\} = \{0.1, 0.2\}$ in the absence of underreporting is $\pmb{\hat{\theta}_{ML}} = \{\hat{\alpha}, \hat{\mu}\} = \{0.100, 0.202\}$ with $\pmb{bias}_{(\pmb{\hat{\theta}_{ML}})} = \{0.000, 0.002\}$. Complete details, including histograms of estimates across trials from each of the five random restarts, appear in Figure S2 in the supporting information. On an interpretable scale, the average estimate of the mean ICU stay duration is $4.95$ days, compared to the ground truth of $5$ days. In the presence of underreporting, the average MLE is $\pmb{\hat{\theta}_{ML}} = \{\hat{\alpha}, \hat{\mu}\} = \{0.142, 0.320\}$ with $\pmb{bias}_{(\pmb{\hat{\theta}_{ML}})} = \{0.042 , 0.120\}$, suggesting that inference in the presence of underreporting remains reasonable but exhibits slight bias compared to the setting where data are simulated exactly from the model. 

\begin{figure}[h]
\centering
\includegraphics[width=.65\linewidth]{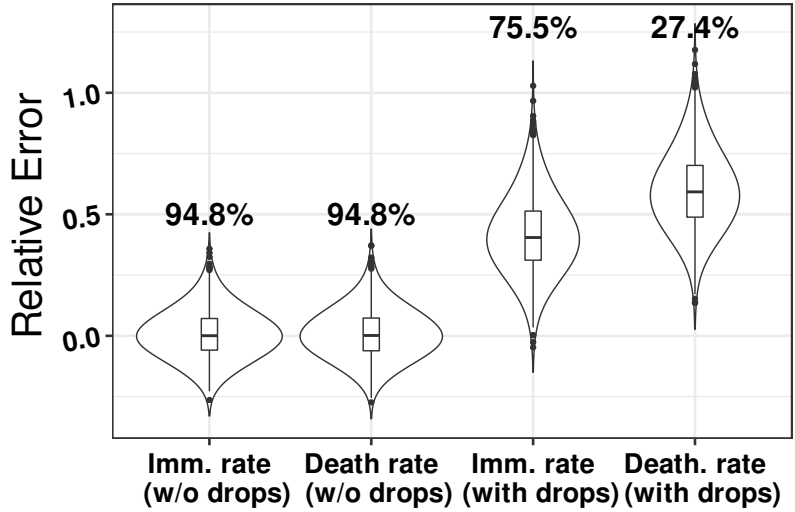}
\caption{Violin plots of relative errors in estimated immigration and clearance rates from the simulation study. The corresponding coverage rates of the $95\%$ confidence intervals are mentioned above the violin plots.}
\label{fig3}
\end{figure}

For a closer look, Figure \ref{fig3} not only represents the violin plots of relative errors in the MLE in both the cases but also shows the coverage of $95\%$ confidence intervals for $\pmb{\hat{\theta}_{ML}} = \{\hat{\alpha}, \hat{\mu}\}$ both in the absence and in the presence of underreporting. Under the true model, the confidence intervals of $(\hat{\alpha})$ and $(\hat{\mu})$ have a coverage of $94.8 \%$ for both parameters. In the misspecified setting with significant underreporting, the coverage of confidence intervals of $(\hat{\alpha})$ and $(\hat{\mu})$ drops noticeably to $75.5 \%$ and $27.4 \%$ respectively. This suggests that while estimates remain in a reasonable range, results and uncertainty estimates especially pertaining to $\mu$ must be interpreted conservatively when only coarse population level data are available.

\subsection{Application to COVID-19 hospitalization data from the University of California, Irvine}
\label{University of California, Irvine}

We now analyze hospitalization data provided by UCI Health, the University of California Irvine's hospital system, collected from August 02, 2020 to December 31, 2020. UCI Health provides us access to anonymized patient-level data which includes the duration of ICU stay for each patient that was diagnosed with COVID-19 and admitted at UCI Health. The availability of such ``line-list" is a rare exception, and in this case study will serve as a ground truth to validate our method, allowing us to check against inference from aggregating the individualized data.  We explore a number of candidate models with variables of interest including the number of hospital beds and ICU beds occupied by patients who have a confirmed diagnosis for COVID-19 and the daily cumulative test positivity rate. Model 3, with
$$\log{\alpha(t)} = \log{\alpha} + \beta_{H} \log{H_{O}(t)} + \beta_{P}P_{A}(t), \quad \mu(t) = \mu,$$
was found to be the best model fit for the UCI Health data via AIC. Table \ref{tab2} gives the maximum likelihood estimates (and 95\% confidence intervals) for all parameters of this model. The estimated clearance rate $(\hat{\mu})$ was $0.08$ $[0.06, 0.11]$ which implied that the average per-patient ICU stay at UCI Health was $12.5$ $[9.1, 16.7]$ days between August 2020 and December 2020. \par
Directly estimating the average ICU stay using the patient-level data by taking the mean directly yields an average per-patient ICU stay at UCI Health is $9.3$ days. Taking this estimate as a gold standard, we find that it is consistent with our results, falling within the $95\%$ confidence interval, $[9.1, 16.7]$ days estimated using our framework. The agreement of our estimate computed from only the aggregated counts with the estimate from the higher-resolution view provided by UCI Health is encouraging, supporting the validity of our proposed approach.

\subsection{Application to COVID-19 hospitalization data from Orange County (OC), CA}
\label{Application to COVID-19 hospitalization data from Orange County, CA}

We analyze hospitalization data from March 29, 2020 to November 15, 2020 from OC, CA. The data consists of the daily total number of hospital beds and ICU beds occupied by patients that have a confirmed COVID-19 diagnosis. Plots of daily hospital and ICU beds occupied in OC revealed distinct trends for three periods in time, namely April 1, 2020 through June 15, 2020, June 15, 2020 through September 1, 2020 and September 2, 2020 to November 15, 2020 which also includes the summer surge in hospital occupancy. The daily cumulative test positivity rates in OC show a similar trend to the number of hospital and ICU beds occupied. The raw data are visualized in Figure \ref{fig:4}.

\begin{table}[h]
\centering
\begin{tabular}{p{8mm} p{18mm} p{20mm} p{20mm} p{20mm}}
Data  & $\hat{\alpha}$ & $\hat{\mu}$ & $\hat{\beta_{H}}$ & $\hat{\beta_{P}}$ \\ \hline
UCI & $0.28$ $[0.08, 0.48]$ & $0.08$ $[0.06, 0.11]$ & $0.13$ $[0, 0.44]$ & $13.00$ $[6.11,19.89]$\\ \hline
OC  & $1.08$ \newline $[0.82,1.42]$ & $0.19$ $[0.16,0.23]$ & $0.49$ $[0.44,0.55]$ & $6.97$ $[5.70,8.24]$\\ \hline
\end{tabular}
\vspace{2mm}
\caption{Maximum likelihood estimates and $95\%$ confidence intervals of the time-homogeneous models that had the lowest AIC values when fitted to the hospitalization data from UCI Health and OC respectively. Model $3$, with $\log{\alpha(t)} = \log{\alpha} + \beta_{H} \log{H_{O}(t)} + \beta_{P}P_{A}(t)$ and $\mu(t) = \mu$ was the best model for both the UCI Health data and OC data.}
\label{tab2}
\end{table}

Since the COVID-19 pandemic was rapidly evolving in OC and the plot of daily total number of hospital and ICU beds occupied by patients showed distinct trends over the entire observation period, we decided to divide the data into three phases. The chosen phases also overlapped with public policy changes in the state of California that may have impacted the number of hospital beds and ICU beds occupied in OC. Phase 1 was chosen from March 29, 2020 to June 15, 2020 while phases 2 and 3 were chosen from June 16, 2020 to September 1, 2020 and from September 02, 2020 until November 15, 2020 respectively. It was on June 18, 2020\footnote{https://www.latimes.com/california/story/2020-06-18/california-mandatory-face-masks-statewide-order-coronavirus-gavin-newsom} when the Governor of California formally announced a mask mandate, and early September when various schools and higher education institutions started reopening after briefly closing for the summer\footnote{https://www.ocregister.com/2020/08/31/coronavirus-reopening-of-orange-county-schools-now-delayed-to-sept-22-at-the-earliest/}. \par

After dividing the data into three phases, we apply our immigration-death framework to estimate all parameters, with a focus on interpreting the per-patient ICU stay as the reciprocal of the clearance rate $\mu$. We first consider analysis under a time-homogeneous model, where we assume that the model parameters mentioned in Table \ref{tab1} are constant across all three phases, i.e., from March 29, 2020 to November 15, 2020 and fit all models based on this assumption.\achal{The constant clearance rate assumption may not be realistic for OC data as suggested by our analysis, but provides a useful baseline comparison that illustrates the advantages of our framework in applying beyond the rigid case of a fixed clearance rate throughout the outbreak.} Next, we fit the parametrized models described in Table \ref{tab1} to each phase separately, referring to this case as the time-inhomogeneous model. \par

The results from fitting each model to the data from OC in the time-homogeneous setting suggests that model 3, with 
$$\log{\alpha(t)} = \log{\alpha} + \beta_{H} \log{H_{O}(t)} + \beta_{P}P_{A}(t), \quad \mu(t) = \mu,$$
is preferable in terms of achieving the has the lowest AIC on the county-level data. Table \ref{tab2} gives the maximum likelihood estimates and the 95\% confidence intervals for the best model for OC. \par

The results based on OC data from March 29, 2020 to November 15, 2020 suggest that the average per-patient ICU stay in OC was $5.3$ $[4.35, 6.25]$ days. We notice that all AIC values for the time-inhomogeneous model were found to be significantly lower than those obtained under the time-homogeneous model. Based on the computed AIC values under the time-inhomogeneous setting, model 2 with
$$\log{\alpha(t)} = \log{\alpha} + \beta_{H} \log{H_{O}(t)} + \beta_{H:P}P_{A}(t)\log{H_{O}(t)},$$
attains the lowest AIC value during phase 1, while model 3 with
$$\log{\alpha(t)} = \log{\alpha} + \beta_{H} \log{H_{O}(t)} + \beta_{P}P_{A}(t), \quad \mu(t) = \mu,$$
attains the lowest AIC values during phases $2$ and $3$. Further details of the AIC values under candidate models are reported in Table S2 appearing in the supporting
information. \par

\begin{figure}[h]
\begin{subfigure}{.48\textwidth}
  \centering
  \includegraphics[width=.9\linewidth]{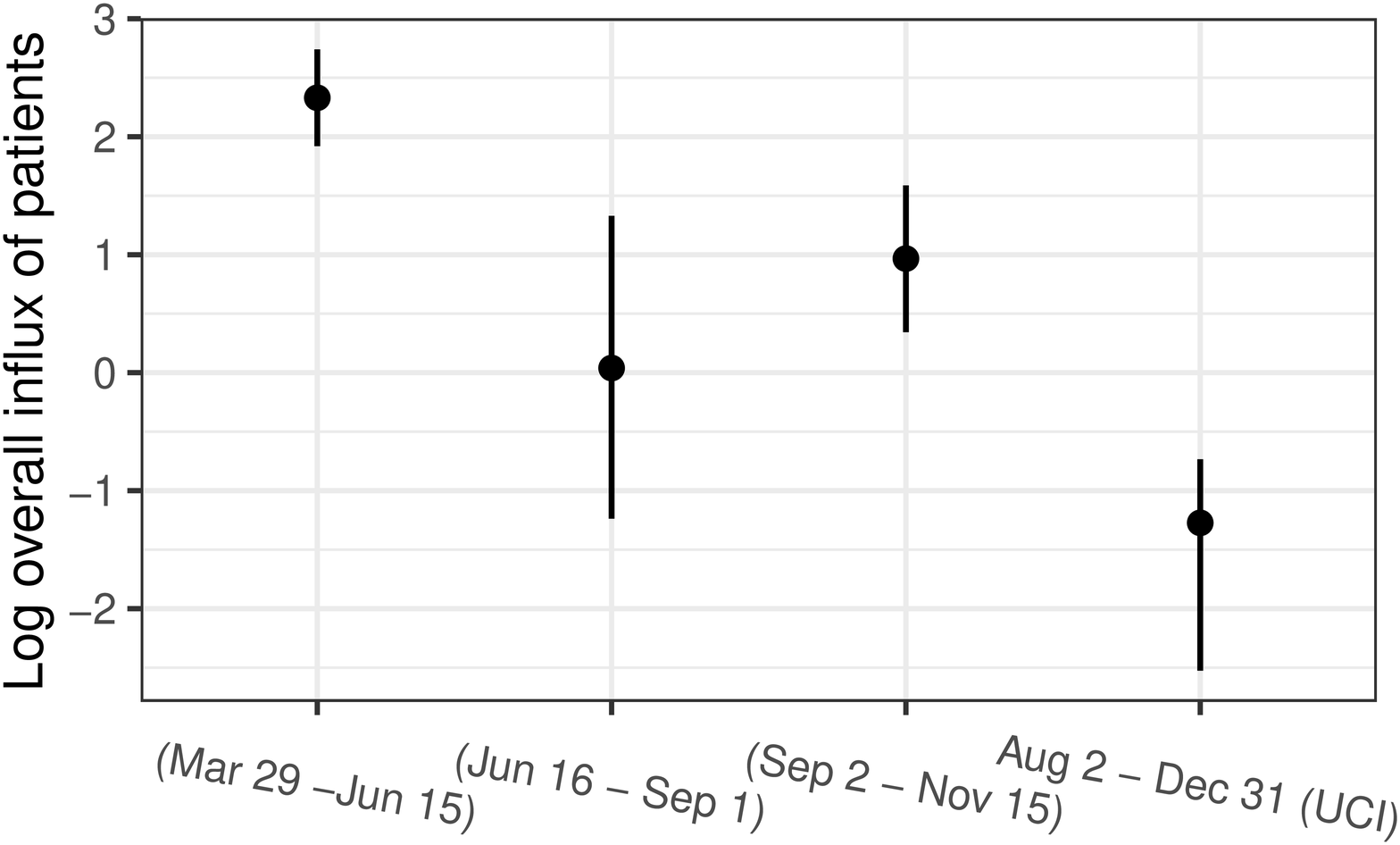}  
  \label{fig5:sub-first}
\end{subfigure}
\begin{subfigure}{.48\textwidth}
  \centering
  \includegraphics[width=.9\linewidth]{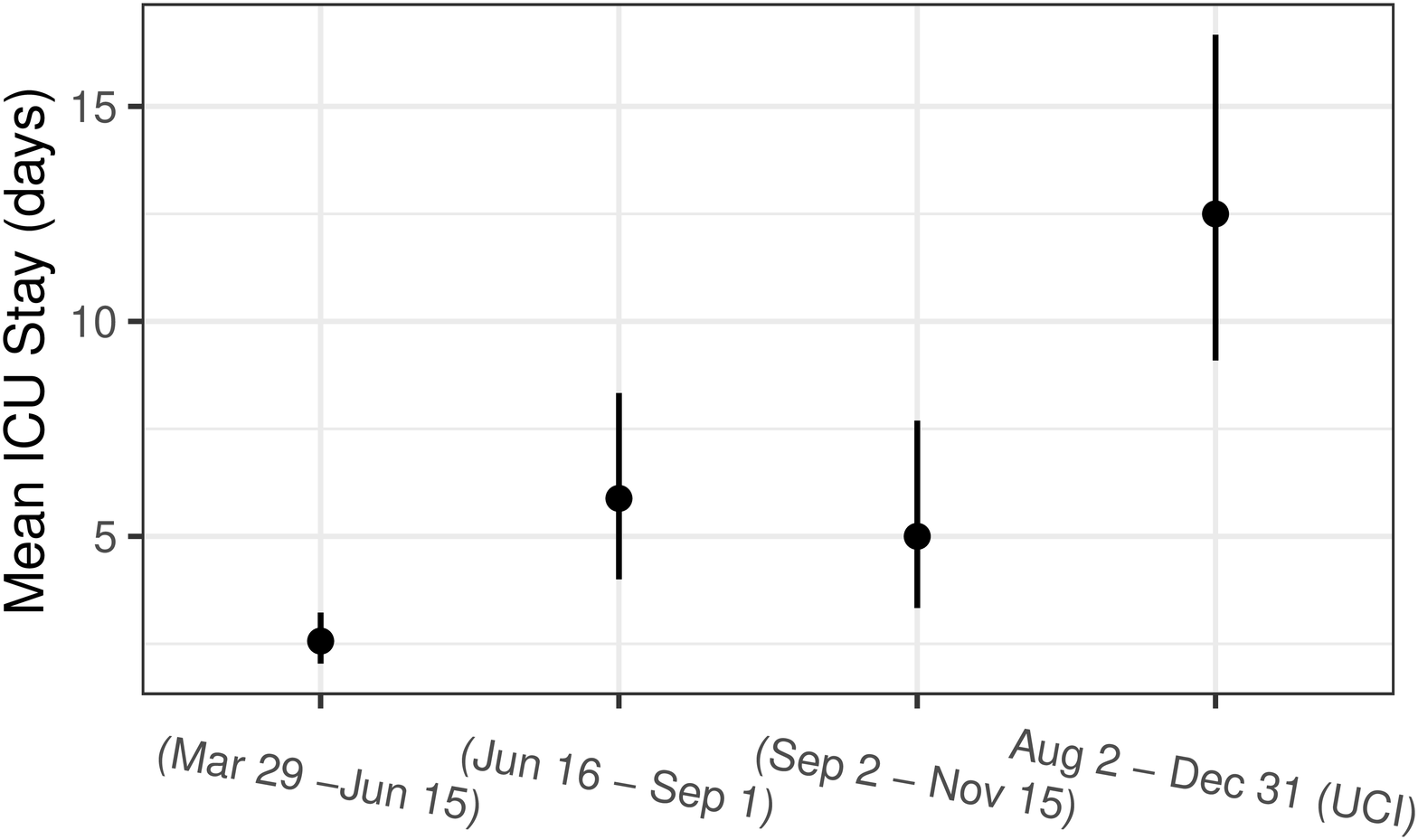}  
  \label{fig5:sub-second}
\end{subfigure}

\begin{subfigure}{.48\textwidth}
  \centering
  \includegraphics[width=.9\linewidth]{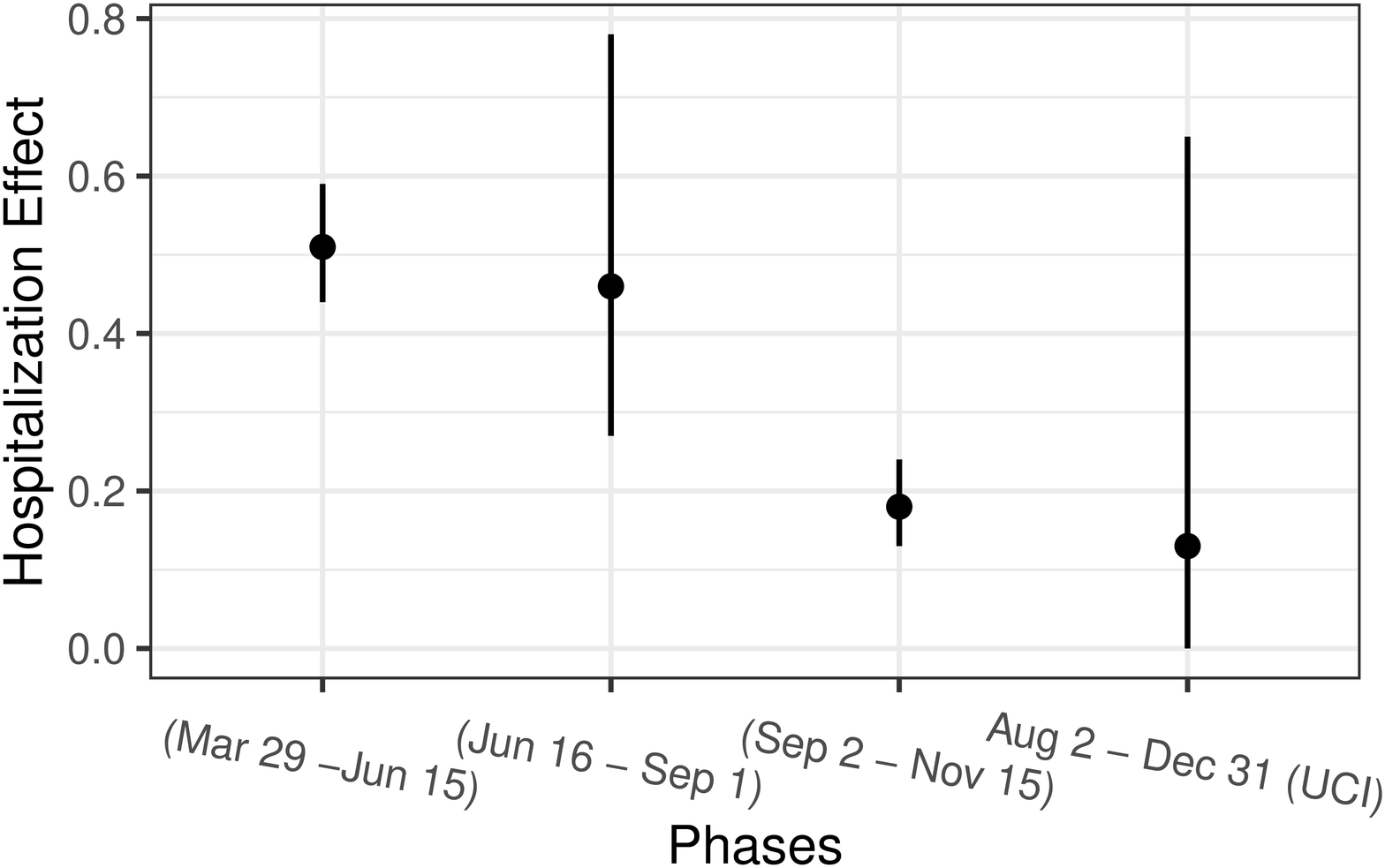}  
  \label{fig5:sub-third}
\end{subfigure}
\begin{subfigure}{.48\textwidth}
  \centering
  \includegraphics[width=.9\linewidth]{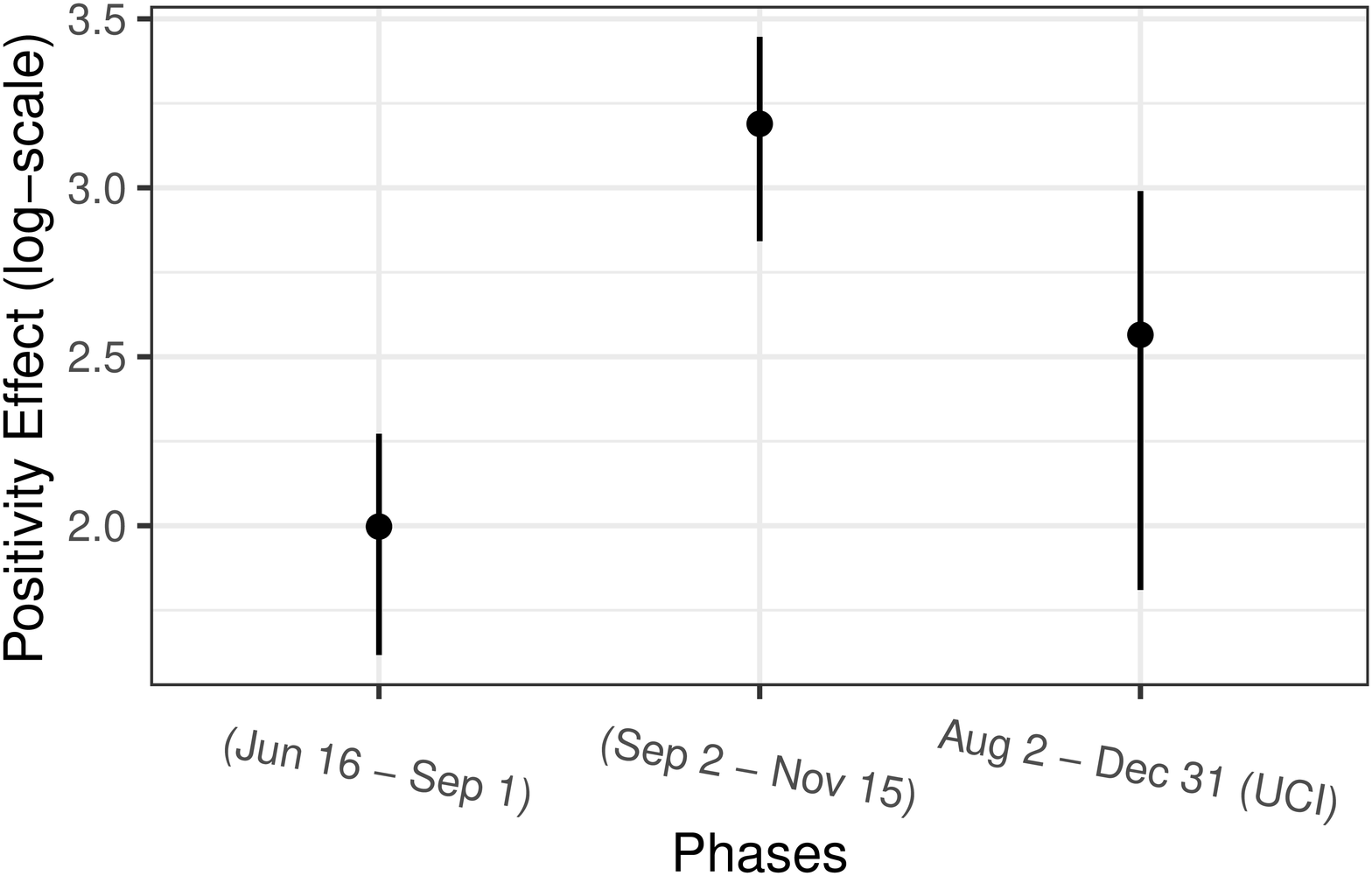}  
  \label{fig5:sub-fourth}
\end{subfigure}
\caption{MLE estimates and 95\% confidence intervals of $\hat{\alpha}, \frac{1}{\hat{\mu}}, \hat{\beta}_{H}, \hat{\beta}_{P}$ respectively, for the best selected model during the three phases in OC.}
\label{fig5}
\end{figure}

During phase 1, our primary parameter of interest, average per-patient ICU stay, was estimated to be $2.56$ days, with 95\% confidence interval $[1.92, 3.45]$. Our inference suggests this average stay increased to $5.99$ $[4.17, 10.0]$ days during phase 2, and then slightly decreased back to $5.00$ $[3.33, 10.0]$ days during phase 3. This is a novel finding from fitting the population level time series, and while there is no validation data such as line-list records available for the county-level study, our validation on the UCI Health data gives us assurance in our results.  Though they are not all as directly interpretable, estimates and confidence intervals for the remaining parameters across all three phases are depicted in Figure \ref{fig5}. We see that phase 2 reflected the largest uncertainty in estimates, while the majority of estimates are statistically significant. A detailed tabulated summary of these results is reported in Table S3 appearing in the supporting information.

\subsection{Model Assessment}

\begin{figure}[h]
\centering
\begin{subfigure}{.49\textwidth}
  \includegraphics[width=.80\linewidth]{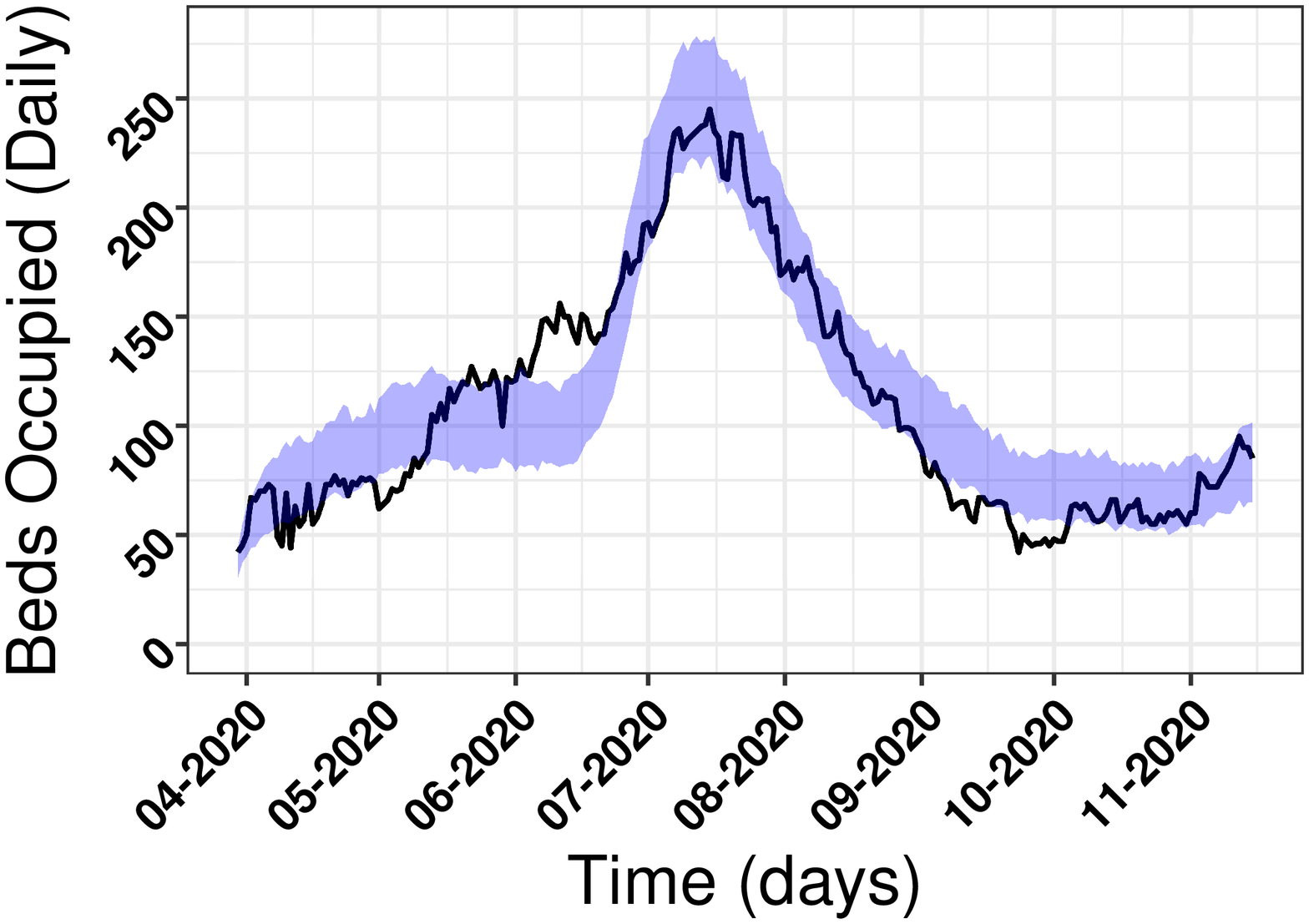}
  \vspace{-0.2cm}
  \caption{}
  \label{fig6:sub-first}
 \end{subfigure}
\begin{subfigure}{.49\textwidth}
  \includegraphics[width=.80\linewidth]{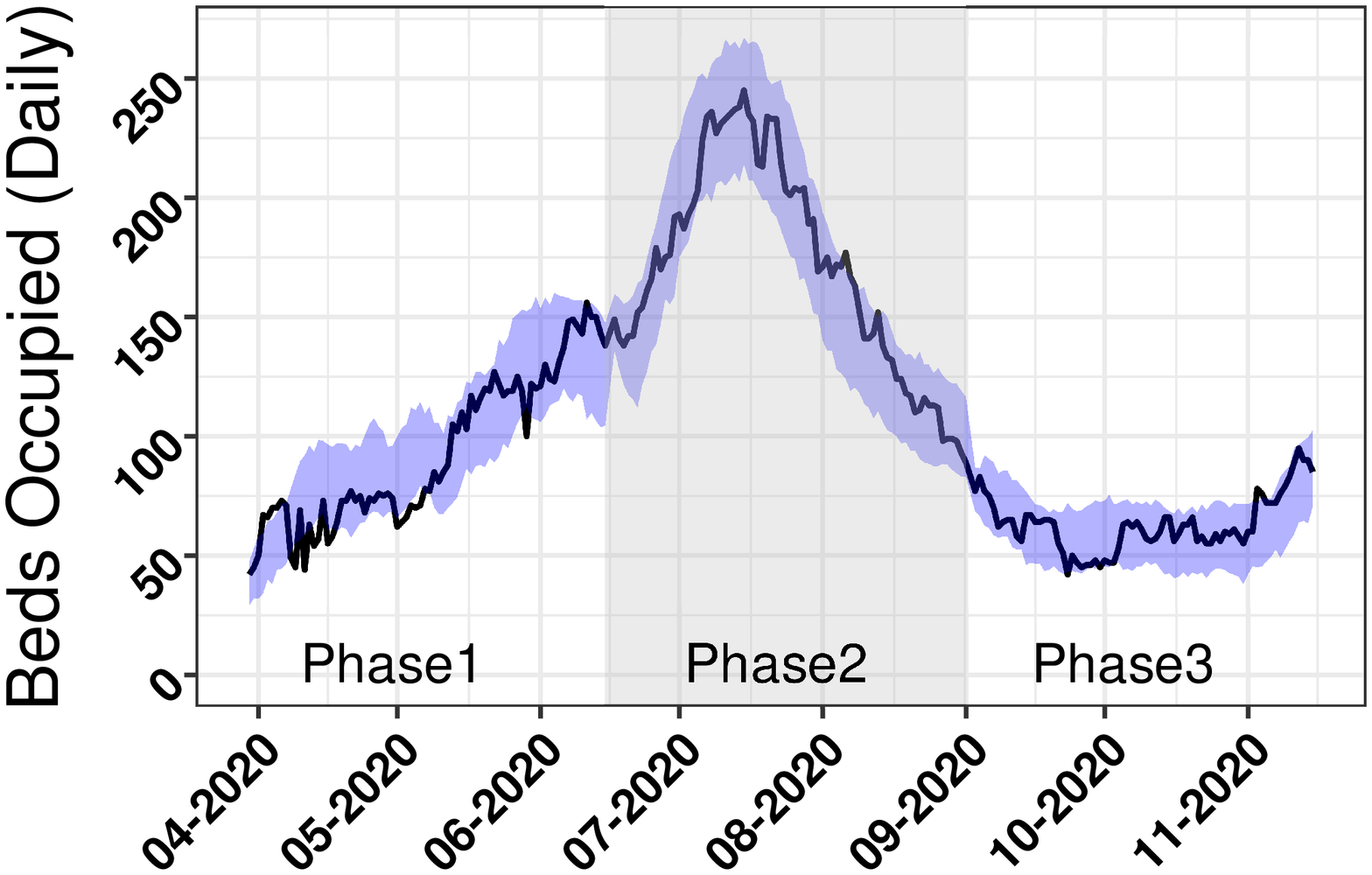}
  \vspace{-0.2cm}
  \caption{}
  \label{fig6:sub-second}
\end{subfigure}
\begin{subfigure}{.49\textwidth}
  \includegraphics[width=.80\linewidth]{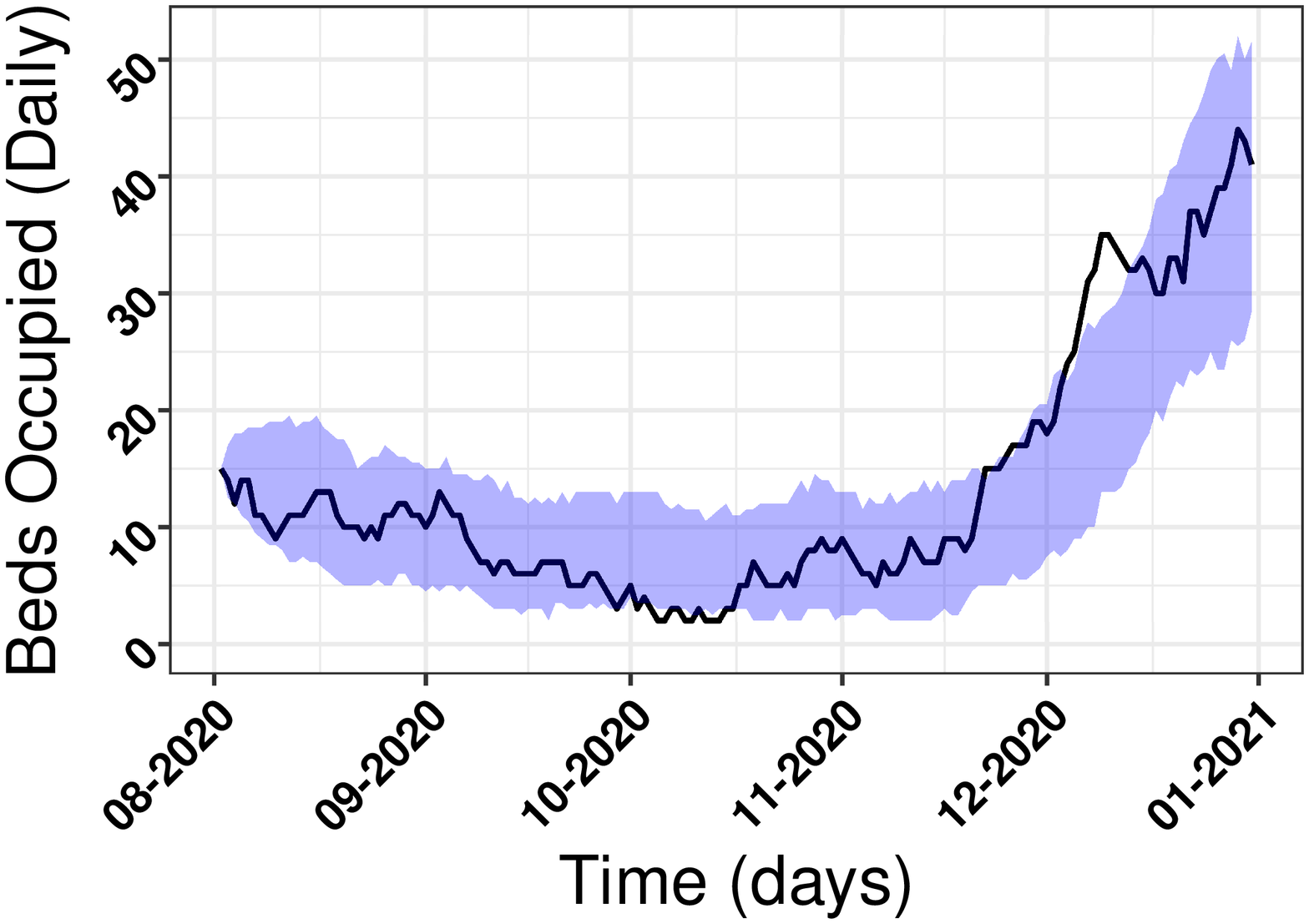}
  \vspace{-0.2cm}
  \caption{}
  \label{fig6:sub-third}
\end{subfigure} 

\caption{\achal{Middle $95\%$ pointwise quantiles of the replicated datasets generated from the best model} for the time-homogeneous model for OC, the time-inhomogeneous model for OC, and the UCI Health system respectively. The black line represents the true data, that is, the daily total number of ICU beds occupied in UCI Health system and OC respectively. Note that the time-homogeneous model for OC cannot capture the true trend as efficiently as the time-inhomogeneous model.}
\label{fig6}
\end{figure}

To further assess the fit of the model, we performed a validation study using a model-based simulation from the estimated parameters of the time-inhomogeneous model,\achal{with the goal being to check and visualize how well our model was able to capture the various aspects of the observed data}. We also evaluated the division of the total time period into three phases by repeating the validation study using the estimated parameters for the time-homogeneous model. For each setting,\achal{we generate 100 replicate datasets consisting of the number of ICU beds occupied conditional on the hospital bed counts}. We then save the $2.5^{th}$ and $97.5^{th}$ percentile of these replicated daily number of ICU beds occupied to\achal{create pointwise middle $95\%$ quantiles generated from these replicated datasets.} \par

Figures \ref{fig6:sub-first}, \ref{fig6:sub-second}, and \ref{fig6:sub-third} display these $95\%$ quantiles for the a) best time-homogeneous model for OC, b) the time-inhomogeneous models for OC, and c) the homogeneous model for UCI Health, respectively. We see that in the time-inhomogeneous case, the predictive interval contains the observed number of beds occupied throughout the time series plot. This suggests that the model successfully balances flexibility with model complexity. On the other hand, this is in contrast to the time-homogeneous model which leads to a much more pronounced discrepancy between the interval\achal{generated from replicated datasets and the observed counts.}

\begin{figure}[ht]
\centering
\begin{subfigure}{.75\textwidth}
\hspace{0.2cm}
  \includegraphics[width=.90\linewidth]{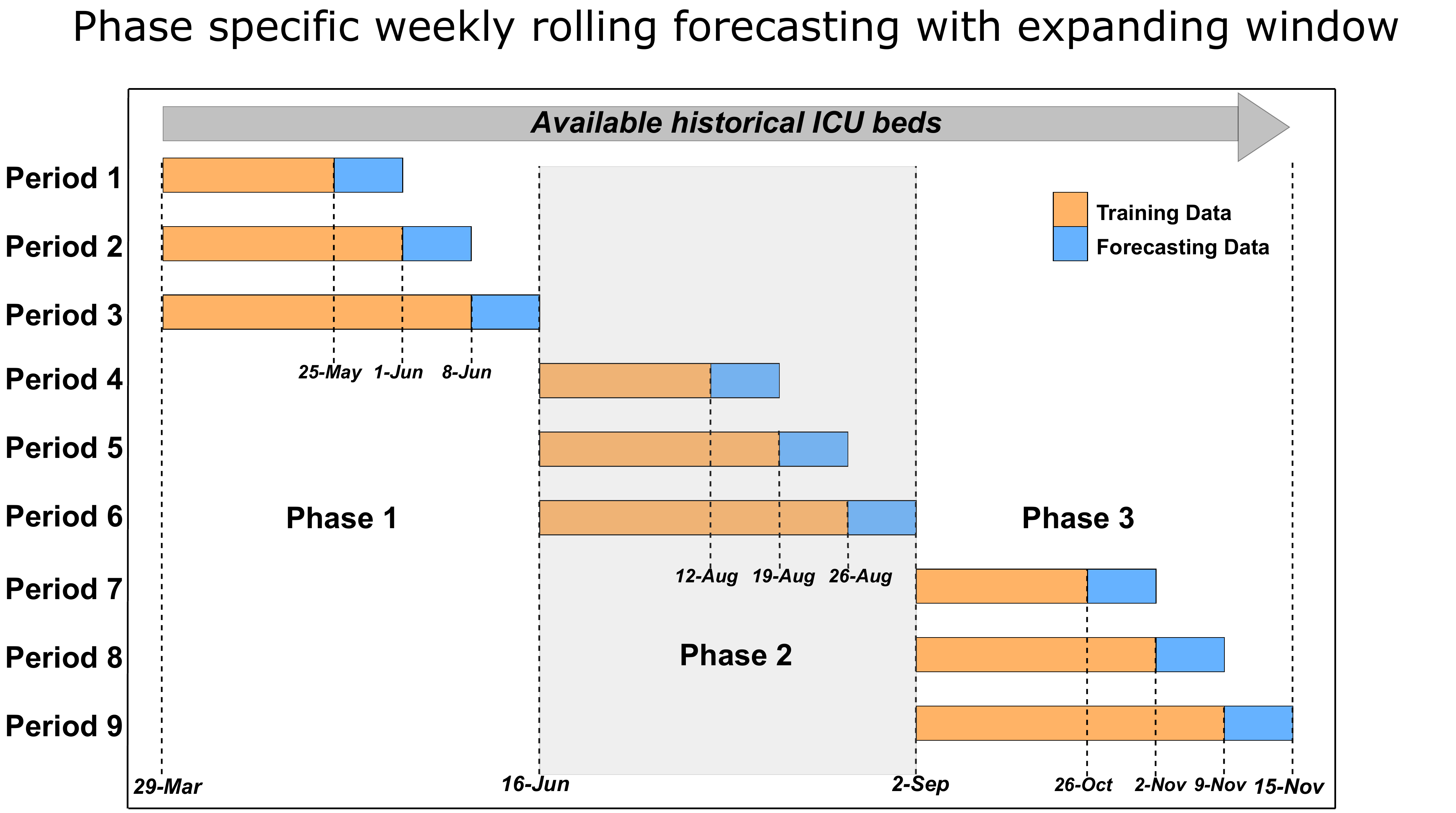}
  \caption{}
  \label{fig7:sub-first}
\end{subfigure}
\begin{subfigure}{.75\textwidth}
  \includegraphics[width=.90\linewidth]{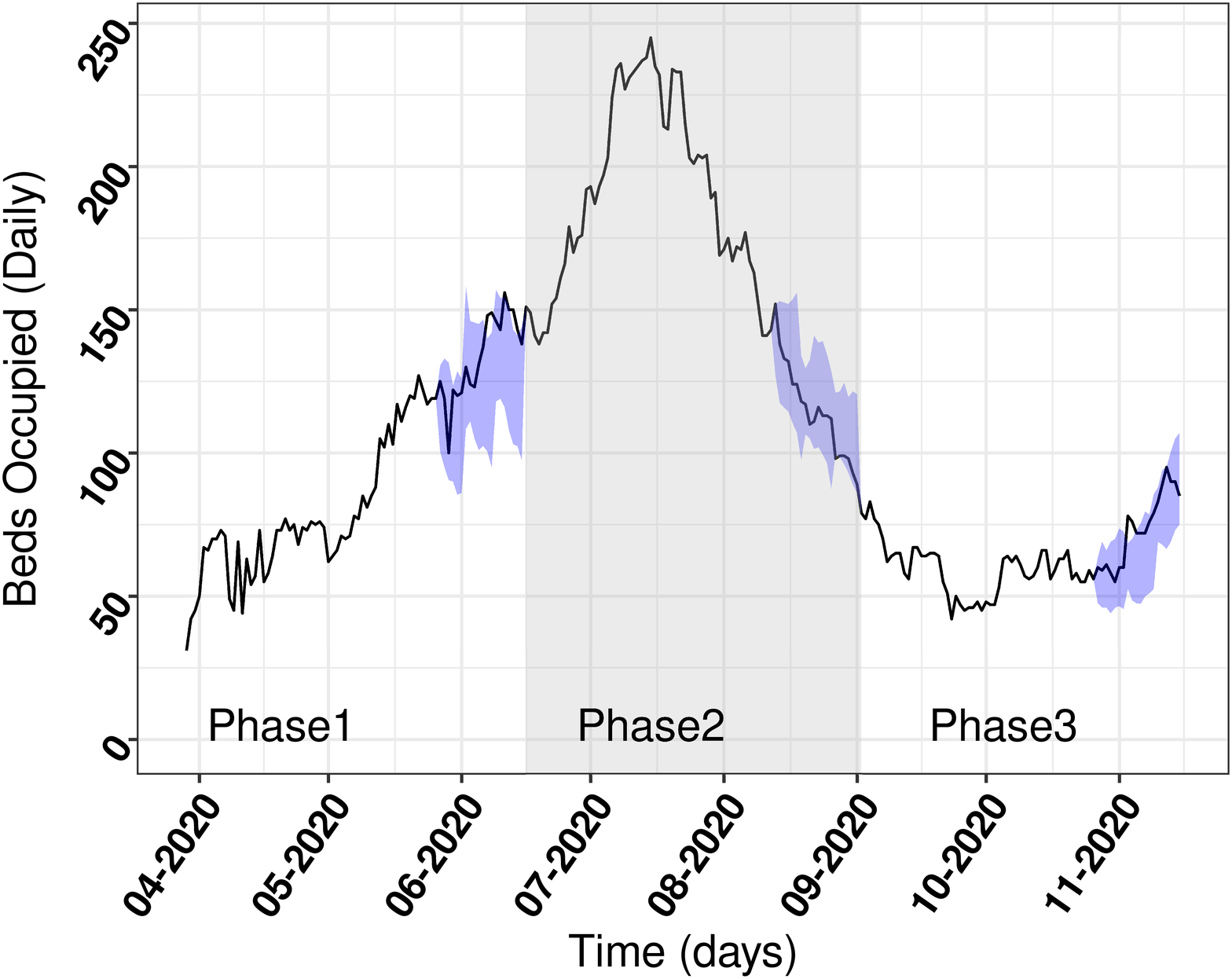}
  \caption{}
  \label{fig7:sub-second}
\end{subfigure}

\caption{\achal{a) Evaluating the predictive performance of our model by implementing a backtesting strategy involving phase specific weekly rolling forecasting with expanding window. 
b) Middle $95\%$ predictive intervals generated from the best model for each phase for OC under the forecasting framework. The black line represents the true data, that is, the daily total number of ICU beds occupied in OC.}}
\label{fig:7}
\end{figure}

\achal{We assess the predictive performance of our model by implementing a \textit{backtesting} strategy involving phase-specific weekly rolling forecasting with an expanding window. To ensure a phase specific expanding training window, each phase is further divided into three distinct periods with Figure \ref{fig2:sub-first} providing a visual representation of the forecasting methodology. This division allows us to progressively include more recent data in each phase, enhancing the model's ability to adapt to temporal variations. By withholding the final three weeks' data for each phase, we establish the initial size of the training window and training data, enabling a comprehensive evaluation of the model's predictive capabilities. 

Specifically, in Phase 1, the training data for Period 1 consists of the number of occupied ICU beds from March 29, 2020, to May 25, 2020. We train the best model specific to Period 1 using this training window and estimate the model parameters. These parameters are then used to generate 100 independent forecasts of the number of occupied ICU beds from May 26, 2020, to June 1, 2020. To quantify the uncertainty associated with these one-week forecasts, we calculate the $2.5^{th}$ and $97.5^{th}$ percentiles, resulting in the middle $95\%$ \textit{predictive} interval.

Upon generating forecasts for the initial training period for Phase 1, we proceed to expand the training window by one week, visually aligning with the training data of Period 2, as indicated in Figure \ref{fig2:sub-first}. Subsequently, the best model for Phase 1 is retrained using this \textit{updated} training window, and the \textit{updated} estimates are utilized to generate a new set of $100$ independent forecasts for the number of occupied ICU beds spanning June 2, 2020, to June 8, 2020 and calculate the corresponding $95\%$ predictive interval. By iteratively following this process, the forecasted number of occupied ICU beds from June 9, 2020, to June 15, 2020 and the corresponding $95\%$ predictive interval are obtained, thus, completing the forecasting of occupied ICU beds for Phase 1.

Following the completion of forecasting occupied ICU beds for Phase 1, a conceptually similar procedure is subsequently implemented for Phases 2 and 3. Particular attention is given to training the best models specific to Phase 2 and Phase 3 when using training data from Periods 4, 5, and 6 and Periods 7, 8, and 9, respectively. Figure \ref{fig2:sub-second} shows the middle $95\%$ predictive interval of the occupied ICU beds computed via phase-specific weekly rolling forecasting with an expanding window. While the $95\%$ predictive interval encompasses the true number of ICU beds occupied across the majority of the time series plot, precedence should be given to the upper limit of the predictive interval when making decisions within a hospital setting.}

\section{Discussion}
\label{s:discuss}

We propose a novel likelihood-based framework to conduct inference on discretely observed data from an immigration-death process that can be adapted to a flexible class of covariate-dependent rates. By deriving the likelihood function for partially observed count data, we are able to numerically perform maximum likelihood estimation and obtain confidence intervals. Moreover, because this framework gives access to criteria such as AIC and BIC, it further allows us to perform model selection among a class of parametric models. By applying our framework, we are able to estimate time-varying parameters describing the per-patient immigration rates into ICUs as well as interpretable durations of ICU stay from population level data collected in OC. Moreover, a similar application restricted to UCI Health data offers validation compared to direct estimates from available line-list data.\achal{COVID-19 was a dynamically evolving pandemic across the United States with studies highlighting key demographic information about patients hospitalized due to COVID-19 and decreasing mortality rates from March, 2020 to November, 2020. \cite{richardson2020presenting, nguyen2021outcomes, roth2021trends} OC was no exception, with hospitalization data revealing distinct trends throughout the observation period, which led us to divide it into three phases. These phases were carefully chosen to correspond to changes in public policy in California.\footnote{https://www.latimes.com/california/story/2020-06-18/california-mandatory-face-masks-statewide-order-coronavirus-gavin-newsom}
\footnote{https://www.ocregister.com/2020/08/31/coronavirus-reopening-of-orange-county-schools-now-delayed-to-sept-22-at-the-earliest/}}\par

\begin{figure}[h]
\begin{subfigure}{0.48\textwidth}
\centering
   \includegraphics[width=0.90\linewidth]{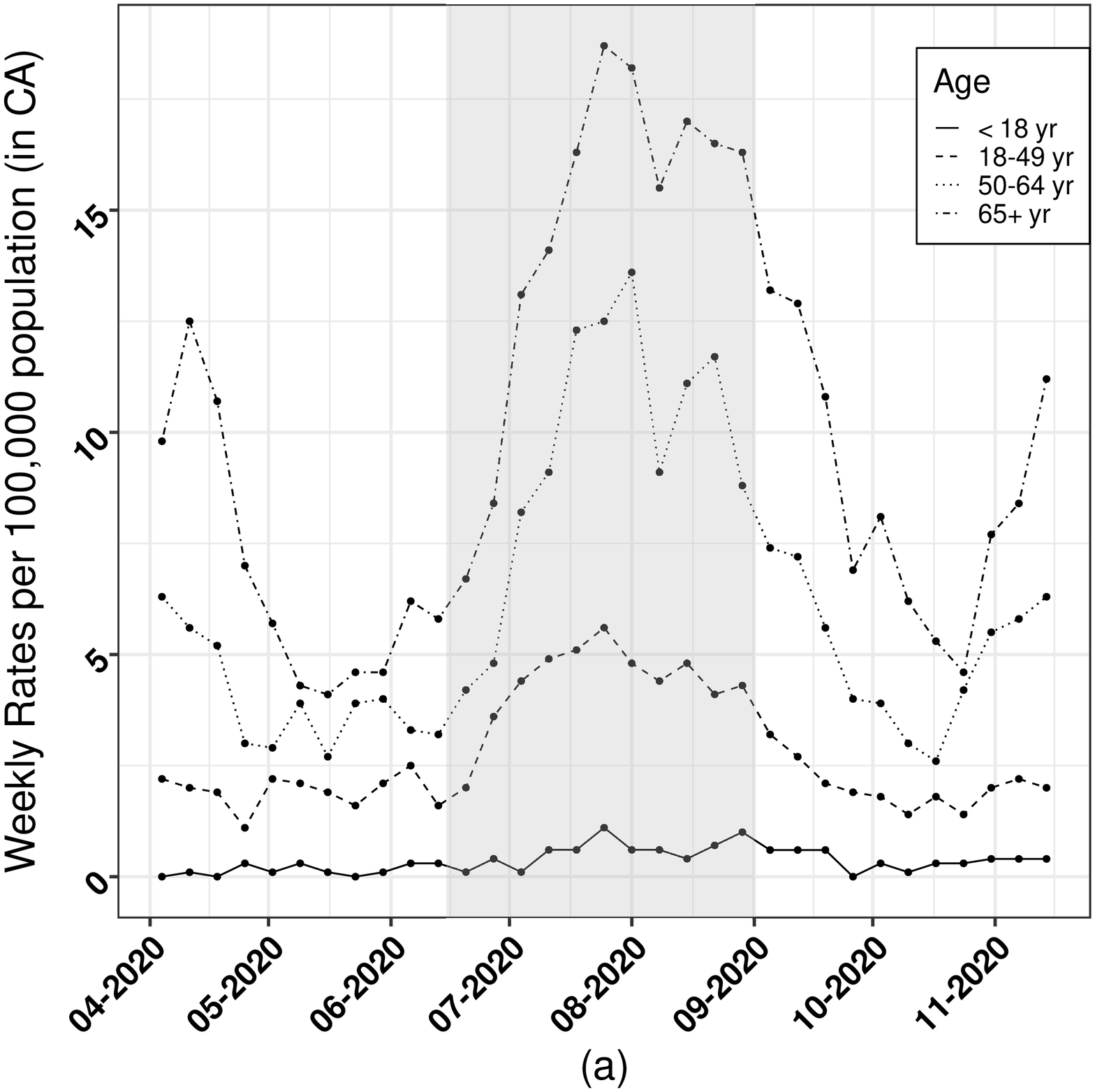}
   \caption{}
   \label{fig:Cali_age}
\end{subfigure}
\begin{subfigure}{0.48\textwidth}
\centering
   \includegraphics[width=0.90\linewidth]{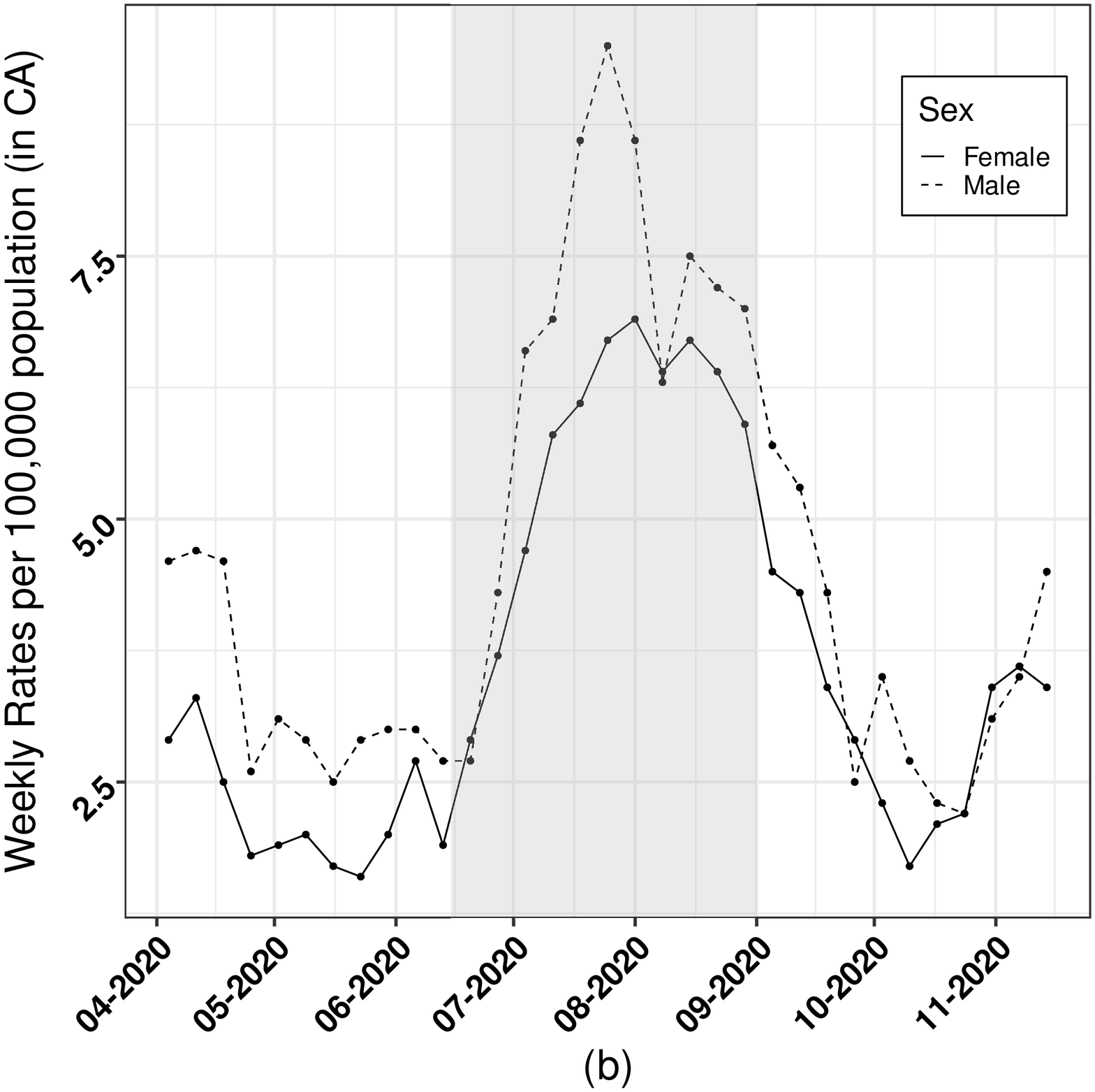}
   \caption{}
   \label{fig:Cali_sex}
\end{subfigure}
\begin{subfigure}{.48\textwidth}
  \centering
  \includegraphics[width=.9\linewidth]{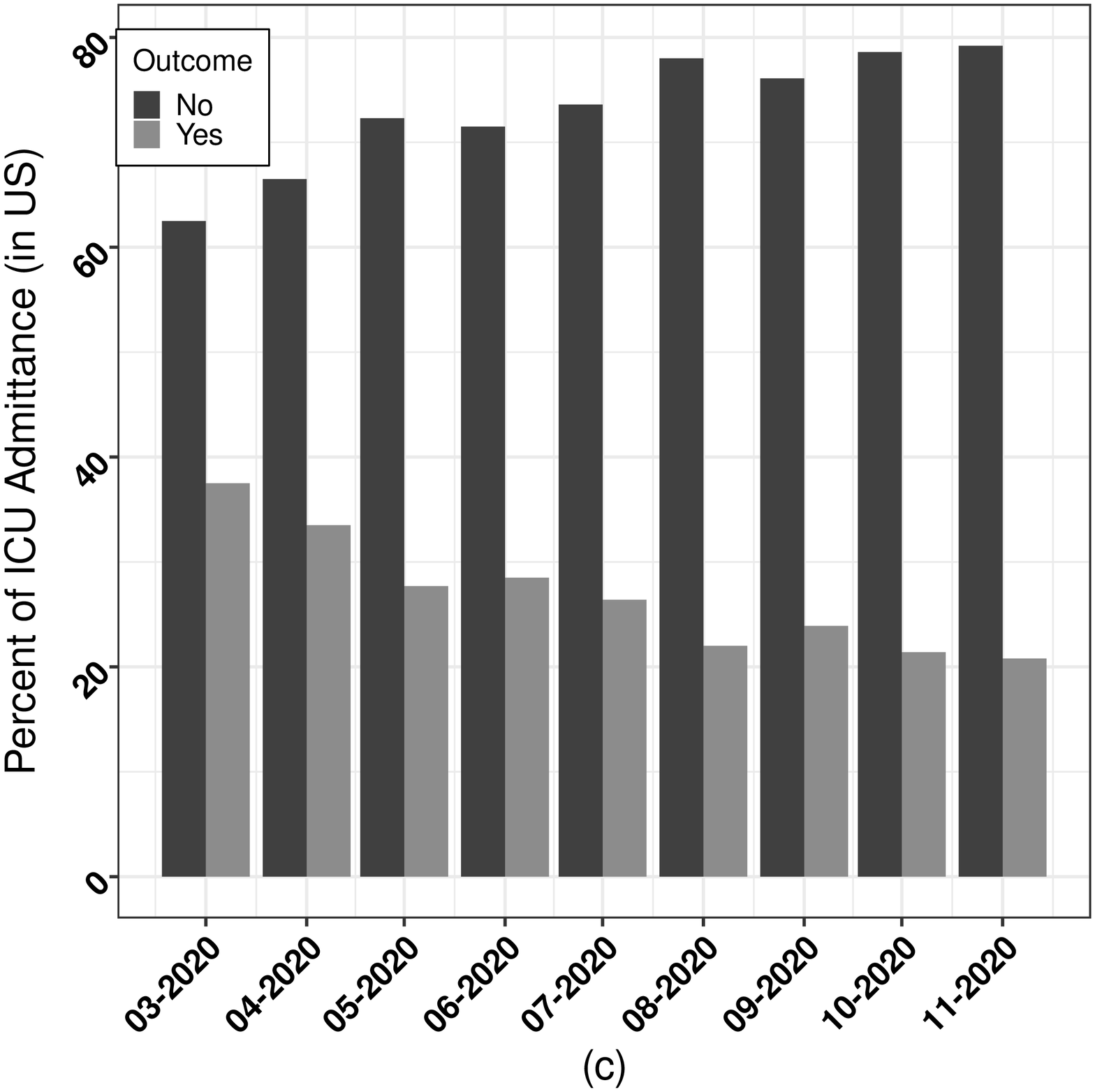}  
  \caption{}
  \label{fig:Cali_ICU}
\end{subfigure}
\begin{subfigure}{.48\textwidth}
  \centering
  \includegraphics[width=.90\linewidth]{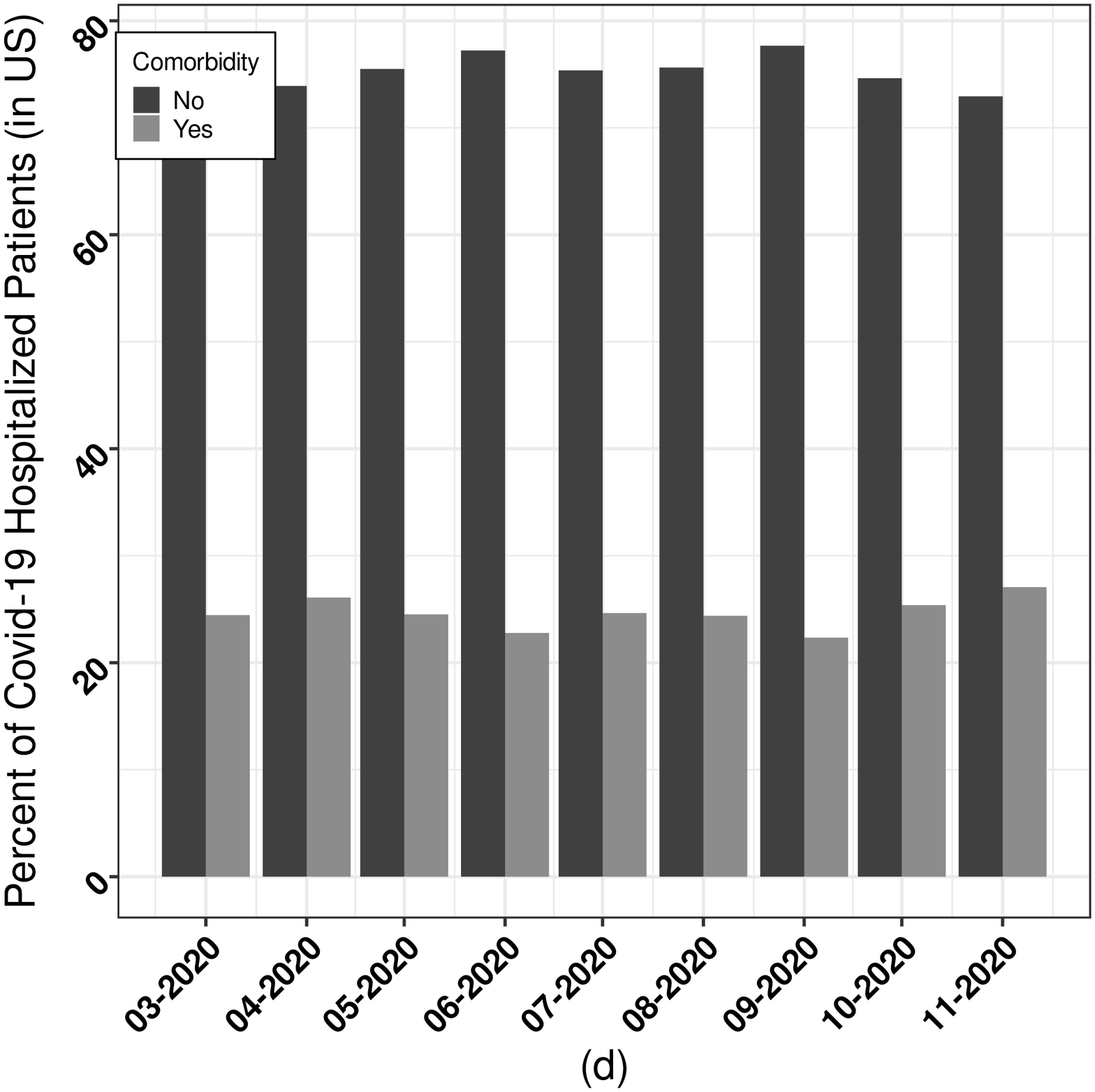}  
  \caption{}
  \label{fig:Cali_comorb}
\end{subfigure}
\caption{Weekly hospitalization rates per $100,000$ population of laboratory confirmed COVID-19 cases in the state of California from March 29, 2020 till November 14, 2020 as provided by The Coronavirus Disease 2019 Associated Hospitalization Surveillance Network (COVID-NET) stratified by (a) age and (b) sex. (c) Percentage of hospitalized patients admitted to the ICU in the United States (d) Percentage of patients hospitalized due to COVID-19 with comorbidities in the United States.}
\label{fig:Cali_demographics}
\end{figure}

\achal{The weekly hospitalization rates in California among people over $65$ years increased during the first half of phase 1 while admission to the ICU among patients hospitalized due to COVID-19 in the US during the months of March and April was relatively higher compared to subsequent months, as shown in Figures \ref{fig:Cali_age} and \ref{fig:Cali_ICU}, respectively. We note the type and location of the data considered in these references vary from the location of the data we consider, with the data from California being the closest to it, so there remains a subjective degree in defining phases to capture general differences in ICU stays broadly. This resonates with a central message of our manuscript about the increased need for granular data availability and transparency of hospital reporting policies during a pandemic.\par

Based on these references, we speculate that the mean ICU stay in OC during phase 1 may have been dominated by precautionary measures taken by hospitals due to the novelty of the disease to routinely shift hospitalized elderly patients to the ICU and due to the high mortality rates during the same period.} At the beginning of the second phase, hospitals were\achal{perhaps} better adapted to handle patients with COVID-19. As a result of the improved guidelines of the state and Federal governments including the formal announcement of a mask mandate by the Governor of California,\achal{hospitals may have become more selective in admitting patients into the ICU which was the case in the United States as suggested by Figure \ref{fig:Cali_ICU}, which may have influenced the mean ICU stay duration.} In general, due to some level of underreporting \achal{and noise} in the data, it is safer to treat our results as slight underestimates of the average per-patient ICU stay. Nonetheless, our validation study using model-based simulation is encouraging, and future work may explore observation models under various emission distributions to directly model the effect of under-reporting. \par

In models developed to forecast hospital occupancy during the COVID-19 pandemic, it is common to obtain parameter estimates using knowledge from other regions. Unlike other mechanistic models, which may calibrate model parameters based on outside information, this paper proposes a methodology for estimating the average per-patient ICU stay using only hospitalization data.\achal{While interval estimates obtained with the I-D model agreed with direct estimates obtained using line list data for UCI hospitalizations, the loss in precision caused by using aggregate data, which was not limited to ignoring risk stratification by hospital, cannot be overlooked. Decision-making on the allocation of hospital ICU resources should always be grounded in the soundest empirical evidence available, and surveillance data may not provide such a foundation for critical decision-making. Hence, we reiterate the need for more granular hospitalization data to be made publicly available during times of outbreak when time is of the essence.}

Together with the mechanistic model's ability to better capture the underlying ICU stay dynamics by allowing the rates to be functions of covariates, the proposed inferential framework can be embedded as extended compartments within models such as SEIR.\cite{bu2021likelihood} Hospitalization and ICU compartments have been key additions within disease models, both because hospitalization data is an observable feature of a pandemic and because of the importance of forecasting hospital system occupancy during an outbreak.\cite{van2020forecasting, morozova2020model} An advantage of our immigration-death framework is that it makes use of surveillance data instead of requiring patient-level data, which is typically difficult to obtain and requires HIPPA compliance. In our current case studies, our model selects daily positivity rate and total number of hospital beds occupied as relevant covariates. Future work may investigate the impact of additional covariates such as the introduction of vaccines and therapeutics, or the implementation of more efficient hospital protocols on ICU admittance rates.\achal{The availability of data at the hospital level may allow us to incorporate covariates that stratify the risk by hospitals and use techniques from survival analysis to extend our model framework.} It would also be fruitful to consider nonparametric estimation of changes in the\achal{clearance} rate, and to examine its effect on the accuracy of the estimated per-patient ICU stay.

\pagebreak

\section*{Acknowledgements}
This work was partially supported by NSF grants DMS 2030355, DMS 2230074, and DMS 1936833, as well as a Donald Bren School of Information and Computer Sciences Exploration Award.

\bibliographystyle{hunsrt}  
\bibliography{icu_stay_manuscript}

\begin{thebibliography}{10}

\bibitem{ferstad2020model}
Johannes~Opsahl Ferstad, Angela~Jessica Gu, Raymond~Ye Lee, Isha Thapa,
  Andrew~Y Shin, Joshua~A Salomon, Peter Glynn, Nigam~H Shah, Arnold Milstein,
  Kevin Schulman, et~al.
\newblock A model to forecast regional demand for {COVID}-19 related hospital
  beds.
\newblock {\em medRxiv}, 2020.
\newblock doi: \url{https://doi.org/10.1101/2020.03.26.20044842}.

\bibitem{ionides2006inference}
Edward~L Ionides, Carles Bret{\'o}, and Aaron~A King.
\newblock Inference for nonlinear dynamical systems.
\newblock {\em Proc Natl Acad Sci USA.}, 103(49):18438--18443, 2006.

\bibitem{breto2018modeling}
Carles Bret{\'o}.
\newblock Modeling and inference for infectious disease dynamics: a
  likelihood-based approach.
\newblock {\em Stat Sci.}, 33(1):57--69, 2018.

\bibitem{doss2013fitting}
Charles~R Doss, Marc~A Suchard, Ian Holmes, Midori Kato-Maeda, and Vladimir~N
  Minin.
\newblock Fitting birth-death processes to panel data with applications to
  bacterial dna fingerprinting.
\newblock {\em Ann Appl Stat.}, 7(4):2315–2335, 2013.

\bibitem{xu2015likelihood}
Jason Xu, Peter Guttorp, Midori Kato-Maeda, and Vladimir~N Minin.
\newblock Likelihood-based inference for discretely observed birth--death-shift
  processes, with applications to evolution of mobile genetic elements.
\newblock {\em Biometrics.}, 71(4):1009--1021, 2015.

\bibitem{crawford2012transition}
Forrest~W Crawford and Marc~A Suchard.
\newblock Transition probabilities for general birth-death processes with
  applications in ecology, genetics, and evolution.
\newblock {\em J Math Biol.}, 65(3):553--580, 2012.

\bibitem{fearnhead2014inference}
Paul Fearnhead, Vasilieos Giagos, and Chris Sherlock.
\newblock Inference for reaction networks using the linear noise approximation.
\newblock {\em Biometrics.}, 70(2):457--466, 2014.

\bibitem{ho2018birth}
Lam Si~Tung Ho, Jason Xu, Forrest~W Crawford, Vladimir~N Minin, and Marc~A
  Suchard.
\newblock Birth/birth-death processes and their computable transition
  probabilities with biological applications.
\newblock {\em J Math Biol.}, 76(4):911--944, 2018.

\bibitem{de2020parameter}
Mathisca de~Gunst, Sophie Hautphenne, Michel Mandjes, and Birgit Sollie.
\newblock Parameter estimation for multivariate population processes: a
  saddlepoint approach.
\newblock {\em Stoch Model.}, 37(1):168--196, 2020.

\bibitem{FintziWakefieldMinin}
Jonathan Fintzi, Jon Wakefield, and Vladimir~N. Minin.
\newblock A linear noise approximation for stochastic epidemic models fit to
  partially observed incidence counts.
\newblock {\em Biometrics.}, 2021.
\newblock doi: \url{https://doi.org/10.1111/biom.13538}.

\bibitem{lescure2021sarilumab}
Fran{\c{c}}ois-Xavier Lescure, Hitoshi Honda, Robert~A Fowler, Jennifer~Sloane
  Lazar, Genming Shi, Peter Wung, Naimish Patel, Owen Hagino, Ignacio~J
  Bazzalo, Marcelo~M Casas, et~al.
\newblock Sarilumab in patients admitted to hospital with severe or critical
  covid-19: a randomised, double-blind, placebo-controlled, phase 3 trial.
\newblock {\em Lancet Respir Med.}, 9(5):522--532, 2021.

\bibitem{KARLIN19751}
Samuel Karlin and Howard~M. Taylor.
\newblock {\em A First Course in Stochastic Processes}.
\newblock 2nd ed. Academic Press, 1975.

\bibitem{lange2003applied}
Kenneth Lange.
\newblock {\em Applied Probability}.
\newblock Springer, New York, NY, 2003.

\bibitem{lange1982calculation}
Kenneth Lange.
\newblock Calculation of the equilibrium distribution for a deleterious gene by
  the finite {F}ourier transform.
\newblock {\em Biometrics.}, 38:79--86, 1982.

\bibitem{NeldMead65}
John~A. Nelder and Roger Mead.
\newblock A simplex method for function minimization.
\newblock {\em Comput J.}, 7:308--313, 1965.

\bibitem{nocedal2006numerical}
Jorge Nocedal and Stephen Wright.
\newblock {\em Numerical Optimization}.
\newblock 2nd ed. Springer Science \& Business Media, 2006.

\bibitem{kochenderfer2019algorithms}
Mykel~J Kochenderfer and Tim~A Wheeler.
\newblock {\em Algorithms for Optimization}.
\newblock MIT Press, 2019.

\bibitem{billingsley1961statistical}
Patrick Billingsley.
\newblock Statistical methods in {M}arkov chains.
\newblock {\em The Ann Math Stat.}, pages 12--40, 1961.

\bibitem{gillespie1977exact}
Daniel~T Gillespie.
\newblock Exact stochastic simulation of coupled chemical reactions.
\newblock {\em J Phys Chem.}, 81(25):2340--2361, 1977.

\bibitem{richardson2020presenting}
Safiya Richardson, Jamie~S Hirsch, Mangala Narasimhan, James~M Crawford, Thomas
  McGinn, Karina~W Davidson, Douglas~P Barnaby, Lance~B Becker, John~D Chelico,
  Stuart~L Cohen, et~al.
\newblock Presenting characteristics, comorbidities, and outcomes among 5700
  patients hospitalized with {COVID}-19 in the {N}ew {Y}ork {C}ity area.
\newblock {\em JAMA}, 323(20):2052--2059, 2020.

\bibitem{nguyen2021outcomes}
Ninh~T Nguyen, Justine Chinn, Jeffry Nahmias, Sarah Yuen, Katharine~A Kirby,
  Sam Hohmann, and Alpesh Amin.
\newblock Outcomes and mortality among adults hospitalized with {COVID}-19 at
  {US} medical centers.
\newblock {\em JAMA Netw. Open}, 4(3):e210417--e210417, 2021.

\bibitem{roth2021trends}
Gregory~A Roth, Sophia Emmons-Bell, Heather~M Alger, Steven~M Bradley,
  Sandeep~R Das, James~A De~Lemos, Emmanuela Gakidou, Mitchell~SV Elkind, Simon
  Hay, Jennifer~L Hall, et~al.
\newblock Trends in patient characteristics and {COVID-19} in-hospital
  mortality in the {U}nited {S}tates during the {COVID}-19 pandemic.
\newblock {\em JAMA Netw. Open}, 4(5):e218828--e218828, 2021.

\bibitem{bu2021likelihood}
Fan Bu, Allison~E Aiello, Alexander Volfovsky, and Jason Xu.
\newblock Likelihood-based inference for partially observed stochastic
  epidemics with individual heterogeneity.
\newblock {\em arXiv}, 2021.
\newblock doi: \url{https://doi.org/10.48550/arXiv.2112.07892}.

\bibitem{van2020forecasting}
Jan-Diederik Van~Wees, Sander Osinga, Martijn van~der Kuip, Michael Tanck,
  M~Hanegraaf, M~Pluymaekers, O~Leeuwenburgh, L~Van~Bijsterveldt, J~Zindler,
  and MT~Van~Furth.
\newblock Forecasting hospitalization and {ICU} rates of the {COVID}-19
  outbreak: An efficient {SEIR} model.
\newblock {\em Bull World Health Organ.}, E-pub: 30 March 2020.
\newblock doi: \url{http://dx.doi.org/10.2471/BLT.20.256743}.

\bibitem{morozova2020model}
Olga Morozova, Zehang~Richard Li, and Forrest~W. Crawford.
\newblock One year of modeling and forecasting {COVID}-19 transmission to
  support policymakers in connecticut.
\newblock {\em Sci Rep.}, 11:20271, 2021.

\end{thebibliography}

\section*{Supporting Information}
The data and open-source code to reproduce all results and experiments described in the article are made available at the authors' webpages, 
while the data for the case study of Orange County is publicly available\footnote{https://occovid19.ochealthinfo.com/coronavirus-in-oc}.

\begin{appendices}
\section{\textbf{Deriving the closed form solution for $\mathbf{\phi_{i}(s,t)}$.}}
\label{Appendix_1}
We can differentiate the PGF to calculate the Kendall's partial differential equation for an immigration-death process (\cite{lange2003applied}). 
On differentiating both sides of equation (\ref{eq2}) in the main text we get,
\begin{align*}
    \frac{\partial}{\partial{t}}\phi_{i}(s,t) &= \frac{d}{dt} \Bigg(\sum_{j=0}^{\infty}\mathbb{P}_{ij}(t) s^{j}\Bigg) \\
    &= \sum_{j=0}^{\infty}\mathbb{P'}_{ij}(t)s^{j}\\
   &= \mathbb{P'}_{i0}(t) +  \sum_{j=1}^{\infty}\mathbb{P'}_{ij}(t)s^{j}\\
    &= -\alpha(t)\mathbb{P}_{i0}(t) + \mu(t) \mathbb{P}_{i1}(t) + \sum_{j=1}^{\infty}\mathbb{P'}_{ij}(t)s^{j} \\
    &= -\alpha(t)\mathbb{P}_{i0}(t) + \mu(t) \mathbb{P}_{i1}(t) \\
    &+ \sum_{j=0}^{\infty}\Big[(-\alpha(t) - j\mu(t))\mathbb{P}_{ij}(t) + \alpha(t)\mathbb{P}_{ij-1}(t) + (j+1) \mu(t) \mathbb{P}_{ij+1}(t)\Big]s^{j} \\
    &= -\alpha(t)\mathbb{P}_{i0}(t) + \mu(t) \mathbb{P}_{i1}(t) - \alpha(t) \sum_{j=1}^{\infty}\mathbb{P}_{ij}(t)s^{j} - \mu(t)\sum_{j=1}^{\infty}j\mathbb{P}_{ij}(t)s^{j} + \alpha(t) \sum_{j=1}^{\infty}\mathbb{P}_{ij-1}(t)s^{j}\\
    &+ \mu(t)\sum_{j=1}^{\infty}(j+1)\mathbb{P}_{ij+1}(t)s^{j} \\
    &= -\alpha(t)\phi_{i}(s,t) - \mu(t) s \frac{\partial}{\partial{s}}\phi_{i}(s,t) + \alpha(t) s \phi_{i}(s,t) + \mu(t) \frac{\partial}{\partial{s}}\phi_{i}(s,t)\\
    &= \Bigg[(1-s)\mu(t) \frac{\partial}{\partial{s}} + (s-1)\alpha(t)\Bigg]\phi_{i}(s,t).
\end{align*}

Thus, the Kendall's partial differential equation for this process is,

\begin{equation}
    \frac{\partial}{\partial{t}}\phi_{i}(s,t) = \Bigg[(1-s)\mu(t) \frac{\partial}{\partial{s}} + (s-1)\alpha(t)\Bigg]\phi_{i}(s,t).
\label{eq3}
\end{equation}

The solution of equation \ref{eq3} gives an explicit expression for the PGF. The solution of equation \ref{eq3} is,
\begin{equation}
    \phi_{i}(s,t) = \exp\Bigg[{{\dfrac{\alpha(t)(1-s)\Big(e^{-\mu(t) t} - 1\Big)}{\mu(t)}}}\Bigg]\Big[1+(s-1)e^{-\mu(t) t}\Big]^{i}.
    \label{eq4}
\end{equation}

\section{\textbf{Algorithm to simulate the number of ICU beds occupied for a set of hospitals over a fixed observation period.}}
\label{Appendix_2}
\begin{algorithm}
\SetAlgoLined
\KwResult{Daily occupied ICU beds}
 initialization $H[0], I[0], T_{max}, \alpha, \mu$\;
 \While{day $< T_{max}$}{
  $\alpha(t) = \alpha H[day]$\;
  $\mu (t) = \mu I[day]$\;
  $\Psi(t) = \alpha(t) + \mu (t)$\;
  wait time $\sim$ Exp($\lambda = \Psi(t))$\;
  \eIf{wait time $<1$}{
  time = time + wait time\;
  $\mathbb{P}(death) = \mu(t)/\Psi(t)$\;
  $\mathbb{P}(immigration) = \alpha(t)/\Psi(t)$\;
  event = sample(immigration or death events with known probabilities)\; 
  \eIf{event = immigration}{
   $I [time] = I [day-1] + 1$\;
   }{
   $I [time] = I [day-1] - 1$\;
  }}
  {$I [time] = I[day-1]$\;
  time = day + 1\;
  day += 1\;
  }
 }
 \caption{Simulation of daily ICU beds occupied in a hospital over a fixed observation period.}
 \label{alg1}
\end{algorithm}

\section{Additional Figures}
\label{Appendix_3}
\setcounter{figure}{0}
\renewcommand{\thefigure}{\Alph{section}\arabic{figure}}

\begin{figure}[H]
    \centering
    \includegraphics[width=12cm]{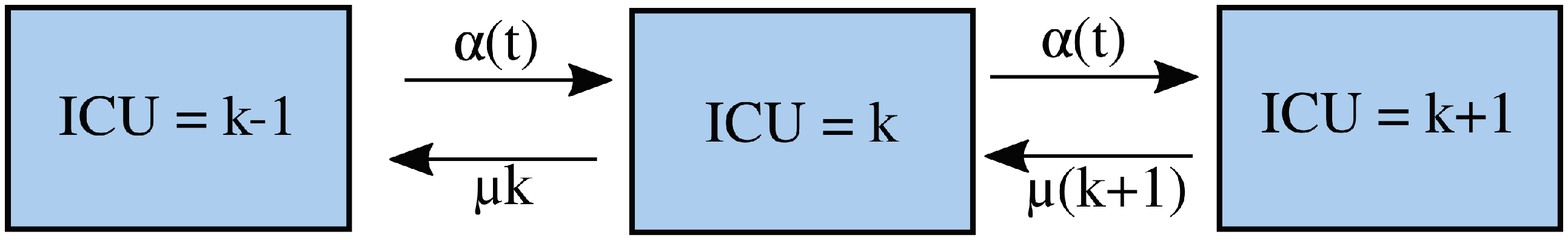}
    \caption{Immigration-death model of ICU capacity. If $k$ beds are occupied in the ICU, then after a random waiting time there can be either $k+1$ or $k-1$ beds occupied. The immigration-death model is characterized by an immigration rate $\alpha(t)$ that does not depend on the current beds occupied in the ICU and a death rate $\mu_k = \mu k$ that is proportional to the current beds occupied in the ICU.}
    \label{fig:A1}
\end{figure}

\begin{figure}[H]
\centering
\subfloat{\includegraphics[width=7cm]{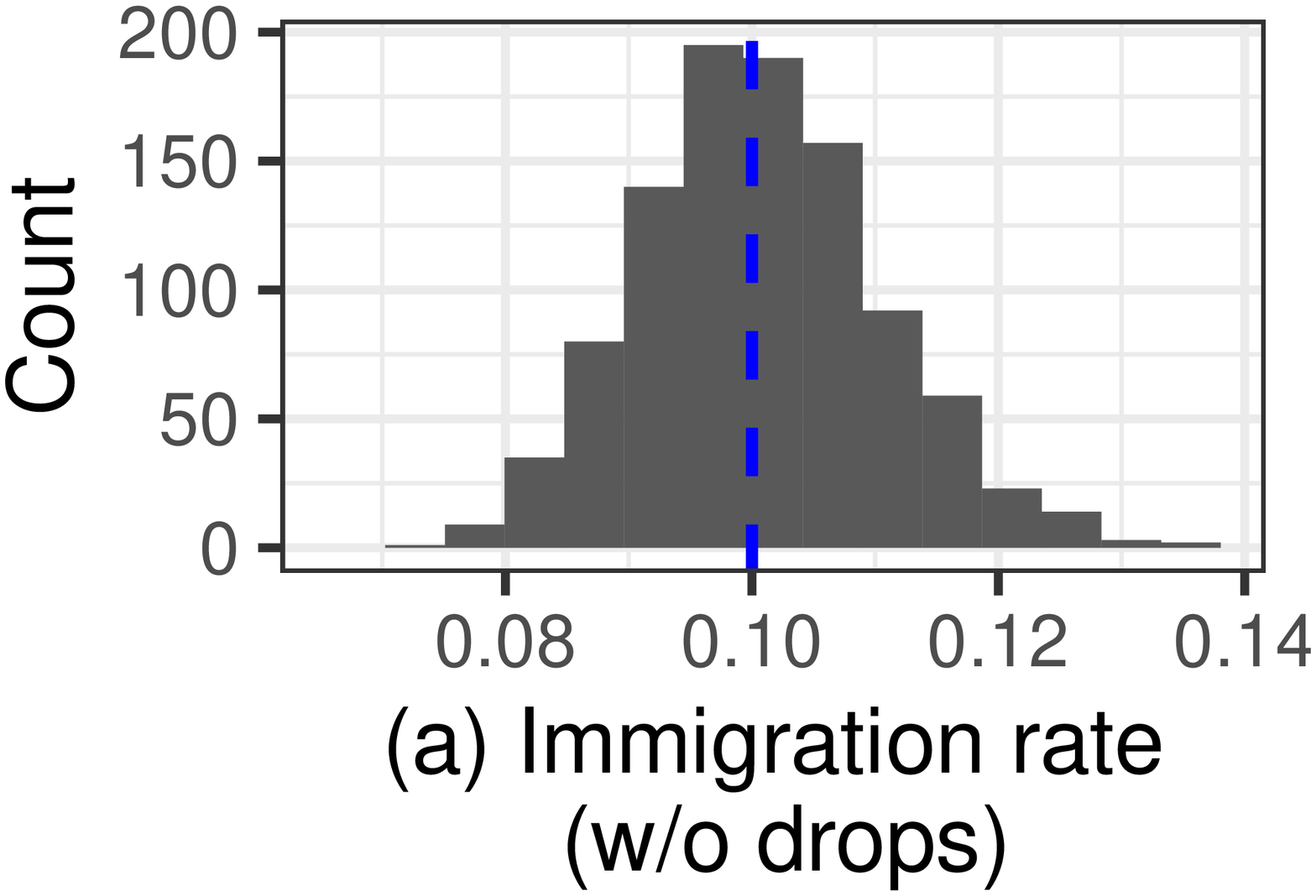}}\hfil
\subfloat{\includegraphics[width=7cm]{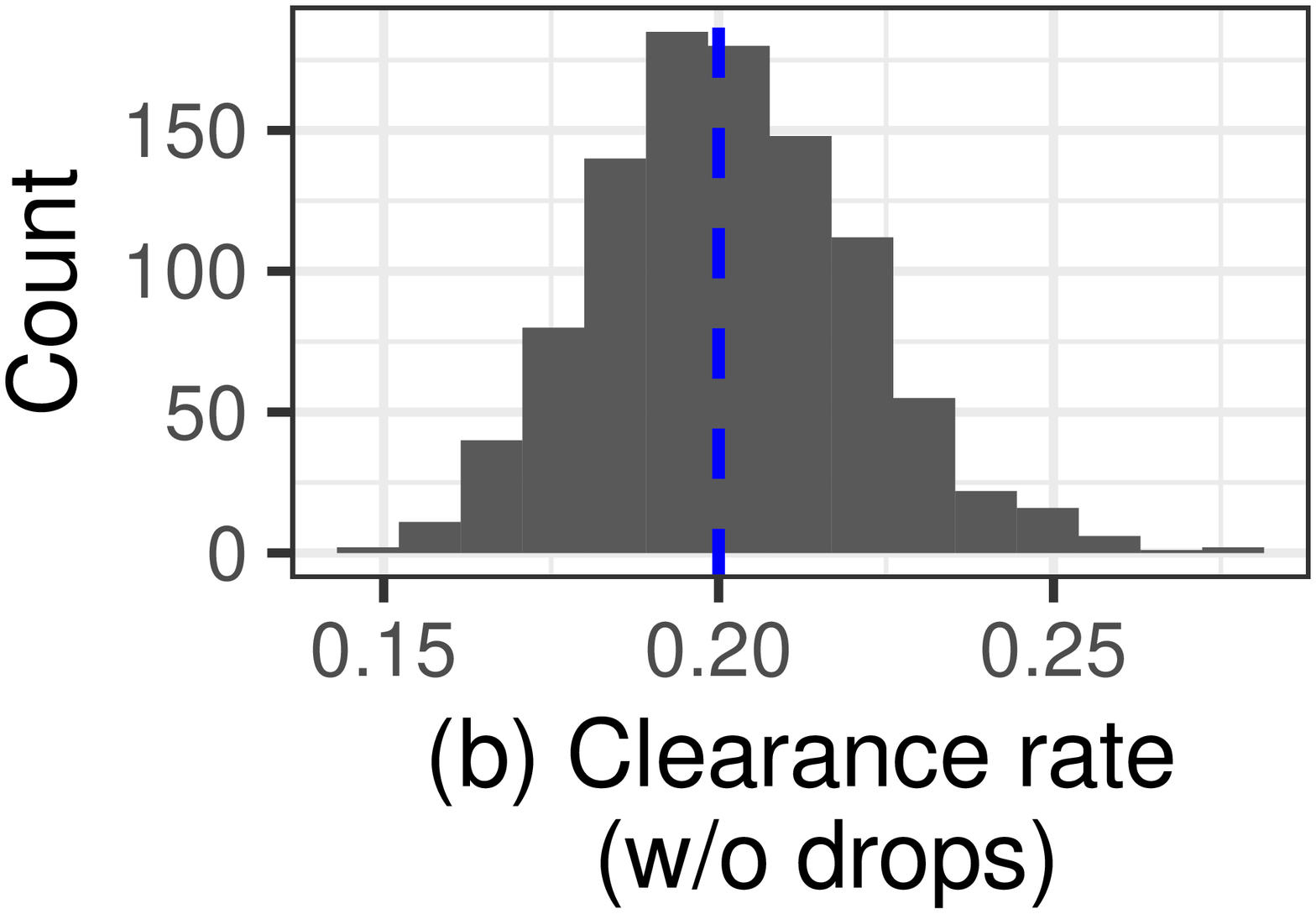}}

\subfloat{\includegraphics[width=7cm]{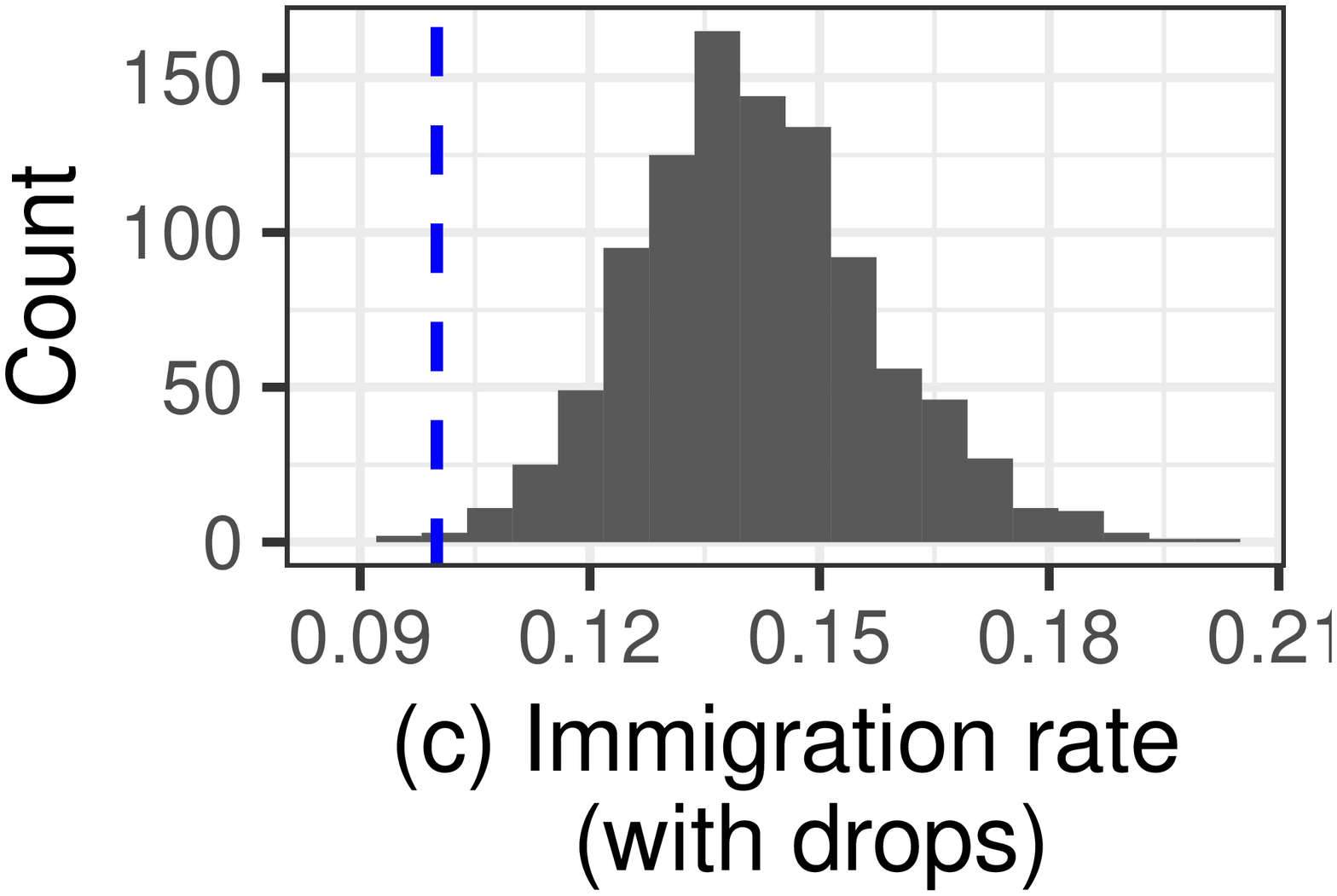}}\hfil   
\subfloat{\includegraphics[width=7cm]{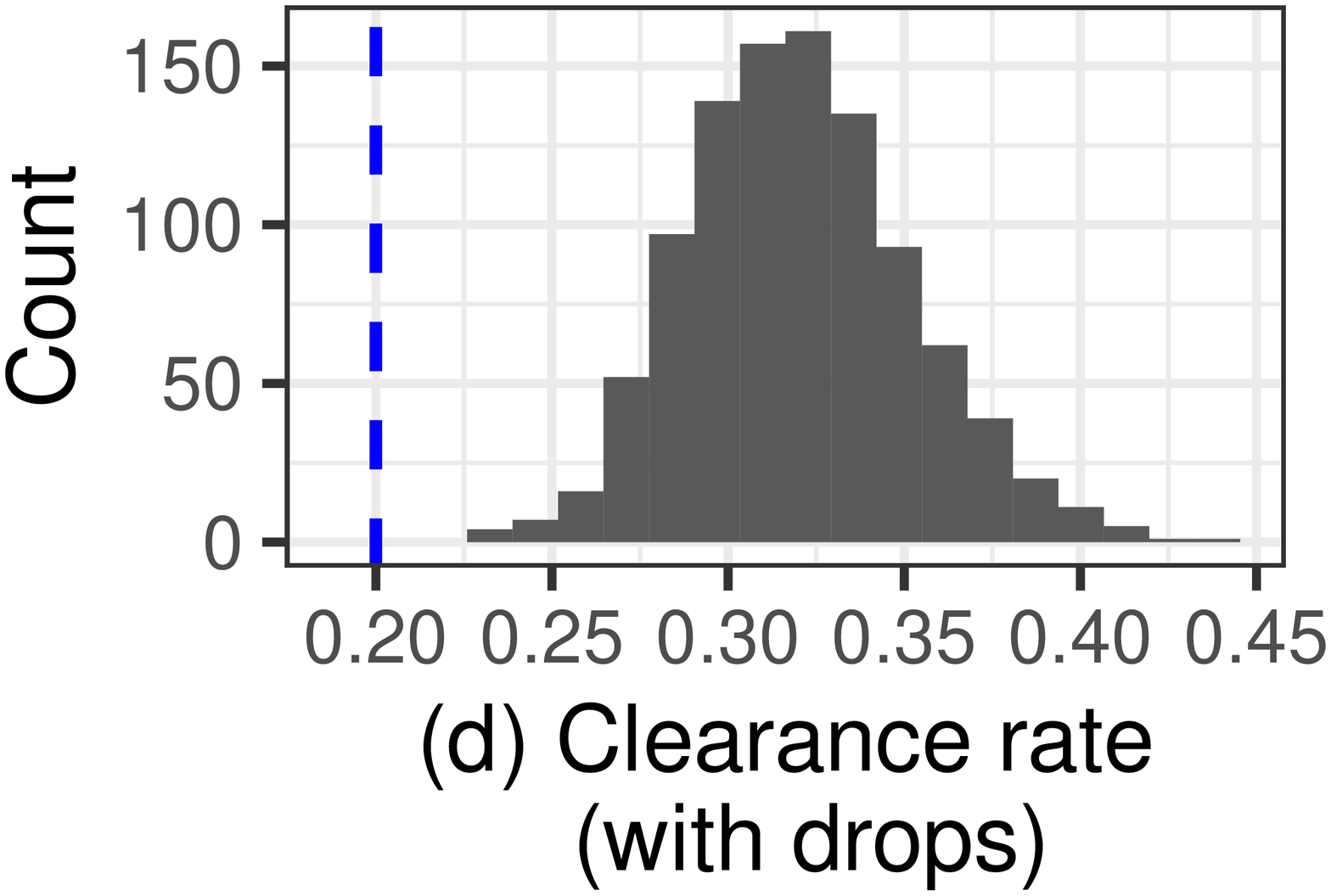}}
\caption{Histograms of estimates of immigration and death rates from the simulation study. (a,c) Histogram of estimates of immigration rates $(\hat{\alpha}$) without any drops and with drops respectively. (b,d) Histogram of estimates of death rates $(\hat{\mu})$ without any drops and with drops respectively. The blue vertical line in each histogram denotes the ground truth.}
\label{fig:A2}
\end{figure}

\section{Additional Tables}
\setcounter{table}{0}
\renewcommand{\thetable}{\Alph{section}\arabic{table}}
\label{Appendix_4}

\begin{table}[h]
\centering
\begin{tabular}{llll}
\textbf{Model} & \textbf{Number of parameters} & \textbf{AIC (UCI)} & \textbf{AIC (OC)} \\ \hline
Model 1 & $5$ & $497.13$ & $1604.37$ \\ \hline
Model 2 & $4$ & $495.01$ & $1601.95$\\ \hline
Model 3 & $4$ & \cellcolor{blue!25}$492.20$ & \cellcolor{blue!25}$1596.72$\\ \hline
Model 4 & $3$ & $498.19$ & $1685.77$\\ \hline
Model 5 & $2$ & $498.06$ & $1724.51$\\ \hline
\end{tabular}
\vspace{2mm}
\caption{AIC values of all time-homogeneous models fitted to the hospitalization data from UCI Health and OC respectively. The lowest AIC values representing the best fitted models are highlighted.}
\label{A1}
\end{table}

\begin{table}[h]
\centering
\begin{tabular}{llllll}
\centering
 & \multicolumn{5}{c}{\textbf{Models}} \\ \hline
\textbf{Phases} & Model 1 & Model 2 & Model 3 & Model 4 & Model 5\\ \hline
Phase 1 & 575.36 & \cellcolor{blue!25}573.49 & 576.53 & 589.25 & 597.07\\ 
Phase 2 & 515.60 & 513.60 & \cellcolor{blue!25}512.90 & 566.98 & 567.84\\ 
Phase 3 & 439.11 & 437.19 & \cellcolor{blue!25}437.03 & 441.96 & 447.45\\
\hline
\end{tabular}
\vspace{2mm}
\caption{AIC values of all time-inhomogeneous models fitted to the hospitalization data from OC. The lowest AIC values representing the best fitted models are highlighted for each phase.}
\label{A3}
\end{table}

\begin{sidewaystable}[h]
\begin{tabular}{lllllll}
\hline
& \multicolumn{6}{l}{\textbf{Orange County (OC)}}\\ \hline
\textbf{Phases} & \textbf{Best Model} & $\hat{\alpha}$ & $\hat{\mu}$ & $\hat{\beta_{H}}$ & $\hat{\beta_{HP}}$ & $\hat{\beta_{P}}$\\ \hline
\textbf{Phase 1} & Model 2 & 10.28 & 0.39 & 0.51 & -3.91 & $-$\\ \hline
95\% confidence interval & & $(6.82, 15.49)$ & $(0.31, 0.49)$ & $(0.44, 0.59)$ & $(-4.83, -2.99)$ & $-$\\ \hline 
\textbf{Phase 2} & Model 3 & 1.04 & 0.17 & 0.46 & $-$ & 7.37 \\ \hline
95\% confidence interval & & (0.29, 3.78) & (0.12,0.25) & (0.27, 0.78) & $-$ & (5.04, 9.70) \\ \hline
\textbf{Phase 3} & Model 3 & 2.63 & 0.2 & 0.18 & $-$ & $24.27$\\ \hline
95\% confidence interval & & (1.41, 4.89) & (0.13, 0.30) & (0.13, 0.24)  & $-$ & $(17.15, 31.39)$\\ \hline
\end{tabular}
\caption{Maximum likelihood estimates and $95\%$ confidence intervals of the time inhomogeneous models that had the lowest AIC values when fitted to the hospitalization data from OC. Model $2$ had the lowest AIC value for phase 1 while model $3$ had the lowest AIC values for phases $2$ and $3$.}
\label{A2}
\end{sidewaystable}

\end{appendices}

\end{document}